%% file: main.tex
\documentclass[sigconf,screen]{acmart}
\usepackage[T1]{fontenc}

\usepackage{pifont}
\usepackage{subcaption}
\usepackage{amsthm}
\usepackage{tikz}
 
\setcopyright{acmcopyright}
\copyrightyear{2024}
\acmYear{2024}
\acmDOI{XXXXXXX.XXXXXXX}

\acmConference[ASPLOS'24]{Architectural Support for Programming Languages and Operating Systems}{April 27--May 01, 2023}{San Diego, CA, USA}

\acmPrice{15.00}
\acmISBN{978-1-4503-XXXX-X/18/06}

\usepackage{multirow}
\usepackage{multicol}
\usepackage{listings}

\definecolor{codegreen}{rgb}{0,0.6,0}
\definecolor{codegray}{rgb}{0.5,0.5,0.5}
\definecolor{codepurple}{rgb}{0.58,0,0.82}
\definecolor{backcolour}{rgb}{0.95,0.95,0.92}

\newcommand{\cmark}{\ding{51}}%
\newcommand{\xmark}{\ding{55}}%

\newcommand*\circled[1]{\raisebox{.4pt}
                    {\tikz[baseline=(char.base)]{
            \node[shape=circle,draw,inner sep=1pt, style={fill=black, text=white}, scale=0.75] (char) {\textbf{#1}};}}}
            
\lstdefinestyle{mystyle}{
    backgroundcolor=\color{backcolour},   
    commentstyle=\color{codegreen},
    keywordstyle=\color{magenta},
    numberstyle=\tiny\color{codegray},
    stringstyle=\color{codepurple},
    basicstyle=\ttfamily\scriptsize,
    breakatwhitespace=false,         
    breaklines=true,                 
    captionpos=b,                    
    keepspaces=true,                 
    numbers=left,                    
    numbersep=5pt,                  
    showspaces=false,                
    showstringspaces=false,
    showtabs=false,                  
    tabsize=2,
}
\lstdefinelanguage{java}{
  keywords={layer, DiffractLayer, Detector, phase_view, prop\_view, to\_3d\_render(), to\_slm(), utils, RingAperture, GaussBeam},
  keywordstyle=\color{blue}\bfseries,
  ndkeywords={class, export, boolean, throw, implements, import, this},
  ndkeywordstyle=\color{darkgray}\bfseries,
  identifierstyle=\color{black},
  sensitive=false,
  comment=[l]{//},
  morecomment=[s]{/*}{*/},
  commentstyle=\color{purple}\ttfamily,
  stringstyle=\color{red}\ttfamily,
  morestring=[b]',
  morestring=[b]"
}

\usepackage{hyperref}

\title{LightRidge: An End-to-end Agile Design Framework for \\ Diffractive Optical Neural Networks} 
 
\author{Yingjie Li}
\email{yingjiel@umd.edu}
\affiliation{%
  \institution{University of Maryland}
  \city{College Park}
  \state{Maryland}
  \country{USA}
  \postcode{20740}
}

\author{Ruiyang Chen}
\author{Minhan Lou}
\author{Berardi Sensale-Rodriguez}
\author{Weilu Gao}
\affiliation{%
  \institution{University of Utah}
  \city{Salt Lake City}
  \state{Utah}
  \country{USA}
  \postcode{84112}
}

\author{Cunxi Yu}
\email{cunxiyu@umd.edu}
\affiliation{%
  \institution{University of Maryland}
  \city{College Park}
  \state{Maryland}
  \country{USA}
  \postcode{20740}
}

\begin{document}

\thispagestyle{plain}
\pagestyle{plain}



\begin{abstract}


To lower the barrier to diffractive optical neural networks (DONNs) design, exploration, and deployment, we propose \textbf{LightRidge}, the first end-to-end optical ML compilation framework, which consists of \textbf{(1)} precise and differentiable optical physics kernels that enable complete explorations of DONNs architectures, \textbf{(2)} optical physics computation kernel acceleration that significantly reduces the runtime cost in training, emulation, and deployment of DONNs, and \textbf{(3)} versatile and flexible optical system modeling and user-friendly domain-specific-language (DSL). As a result, LightRidge framework enables efficient end-to-end design and deployment of DONNs, and significantly reduces the efforts for programming, hardware-software codesign, and chip integration. Our results are experimentally conducted with physical optical systems, where we demonstrate: \textbf{(1)} the optical physics kernels precisely correlated to low-level physics and systems, {\textbf{(2)} significant speedups in runtime with physics-aware emulation workloads compared to the state-of-the-art commercial system,} \textbf{(3)} effective architectural design space exploration verified by the hardware prototype and on-chip integration case study, and \textbf{(4)} novel DONN design principles including successful demonstrations of advanced image classification and image segmentation task using DONNs architecture and topology.

\end{abstract}
 
\maketitle

\input{isca2022-latex-template/01sec-intro}

\input{isca2022-latex-template/03sec-donn_2}

\input{isca2022-latex-template/04sec-compiler}

\input{isca2022-latex-template/05sec-DSE}

\input{isca2022-latex-template/07sec-results}

\input{isca2022-latex-template/10sec-conclusion}

\newpage

 \appendix
 \input{isca2022-latex-template/ae}

\bibliographystyle{IEEEtranS}
\bibliography{bib/Yingjie.bib, refs.bib}

\newpage


\end{document}

%% file: isca2022-latex-template/01sec-intro.tex
\section{Introduction}\label{sec:intro}

Deep neural networks (DNNs) have experienced substantial growth in recent years, making significant contributions in many application domains like autonomous systems, natural language processing, and health care\cite{cao2016deep, covington2016deep, lecun2015deep, szegedy2013deep, erhan2014scalable, fathi2018deep, beam2018big}.
{However, large DNN models producing high system throughput, usually suffer from high carbon footprint. For example, recent studies estimated 626,000 pounds of planet-warming carbon dioxide, equal to the lifetime emissions of five cars, produced in training Transformer network \cite{strubell2019energy,sanh2019distilbert}. On the other side, the embedded accelerators \cite{simon2019blade,cass2019taking,verhelst2017embedded,tambe2021edgebert,reagen2016minerva,wu2019machine}, which are designed to improve resource and power efficiency, suffer from limited functionality and throughput. Thus, while there have recently seen great progress in customized accelerators that adjust the computing performance with efficiency in hardware architectures and systems, the Pareto-frontier of conventional accelerators remains the same \cite{dean20201,lecun2015deep,ragan2013halide,genc2021gemmini,jouppi2018domain,rao2018beyond}.  
}

To advance the Pareto-frontier of ML systems, i.e., offering high computing performance as well as high power efficiency, accelerators taking advantage of optics, namely \textit{optical neural networks} (ONNs), have recently attracted significant interest in machine learning and hardware acceleration. The main advantages of ONNs over digital accelerators can be summarized as follows -- \textbf{(1)} In optical computing systems, since the input features are encoded and carried by light, the computation and data movement will happen at the {speed of light in the medium} with orders of magnitude advantages in computation speed \cite{gu2020towards,shen2017deep,hamerly2019large,xu202111,gao2021artificial,li2021late,li2022physics,chen2022complex}. 
{\textbf{(2)} The laser implemented in the optical systems can be easily expanded with passive optical devices, such as beam splitters, to multiple channels, which means parallel computation can be easily realized with ONN systems, and the throughput of the system will be significantly increased \cite{lin2018all,mengu2020scale,zhou2021large,chen2022physics}. \textbf{(3)} The trained ONN system will be deployed with passive optical devices, which means there is no additional energy cost for all-optical inference process, thus improving the energy efficiency significantly \cite{shen2017deep,ying2020electronic,gu2020towards,feldmann2021parallel,li2021real,qian2020performing,yan2019fourier,zhou2020situ,tang2023device}}. 
\textit{Diffractive Optical Neural Networks} (DONNs) is one of the most promising research areas in ONNs, which mimic the propagation and connectivity properties of conventional neural networks, by utilizing the nature physics of light diffraction and phase modulation of coherent light \cite{lin2018all,li2021real,mengu2020scale,rahman2021ensemble,zhou2021large,chen2022physics,li2022physics}. Even though the inference of the physical DONN is all optical, the training part that leads to its design is done through digital platforms, {where a precise, efficient and hardware-aware emulation engine is required. 

The existing optical emulation engines, such as Mathworks BeamLab \cite{veettikazhy2021bpm} and LightPipes \cite{vdovin1997lightpipes} \footnote{\small LightPipe has been maintained for commercial uses (\url{http://www.okotech.com}) and we compare with the latest version at \url{https://github.com/opticspy/lightpipes}.}, mainly focus on the emulation of the physical phenomenon while lack the key functionalities and domain-specific runtime optimizations in supporting the developments of DONNs. Specifically, it is particularly challenging for existing optical emulation frameworks to deal with DONN training and inference due to the following reasons: (1) The core emulation functions are not differentiable, which makes the backpropagation-based training hard to implement. (2) The implementation is not optimized in runtime. For example, LightPipes does not support tensor representations and operator fusion, which significantly limits the runtime performance (see Table \ref{table:compiler_comp}). (3) There does not exist hardware/device aware emulation supports, which require significant extra efforts for correlating numerical emulations and physical deployments. 
}

The critical technical barriers in design, training, exploration, and hardware deployment of DONNs are summarized as follows:

\noindent
\textbf{Challenge 1:}
{Sufficient multi-disciplinary domain-knowledge in optical physics, fabrication, and machine learning (ML) are required for DONN system design and deployment, which puts a critical technical barrier to exploring and advancing DONN systems in real-world applications. 
At this point, there does not exist an end-to-end design framework that supports design and exploration for full-stack DONNs design, optimization, fabrication, and on-chip integration.
Moreover, the broad architectural search space with software, optics, and fabrication hyperparameters can be an obstacle for efficient design space exploration (DSE), which also motivates the development of an end-to-end design framework.}

\begin{figure}
    \centering
    \includegraphics[width=1\linewidth]{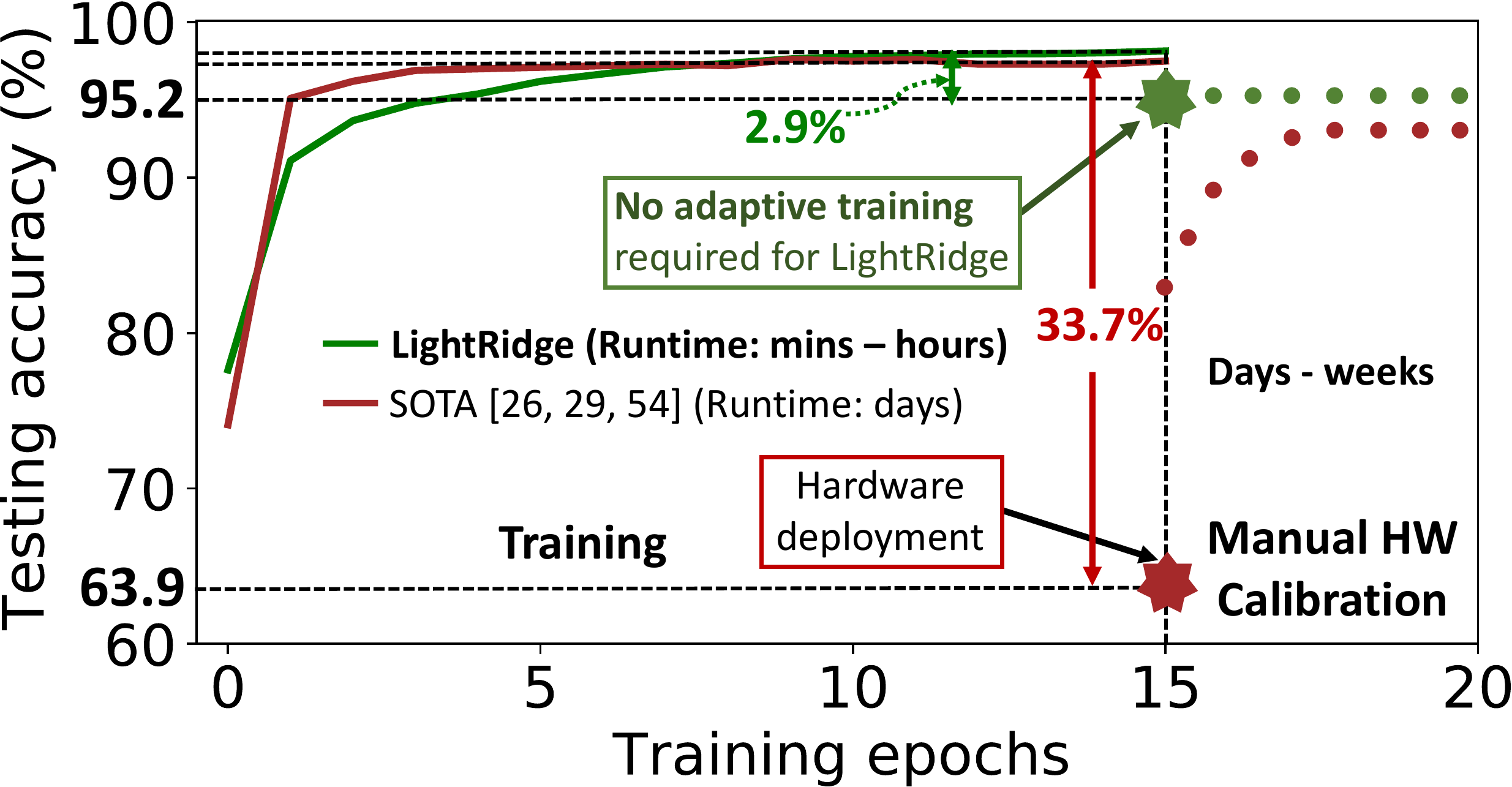}
    \caption{Model performance and time-to-deployment runtime comparison between hardware-in-loop adaptive training with manual calibration and LightRidge -- (1) LightRidge reduces design cycle from days to hours with high-performance emulation kernels and DSE engine; (2) LightRidge results in significantly improved correlation in out-of-box deployment, which gets rid of expensive manual hardware calibration processes.
    }
    \label{fig:intro_correlation}
\end{figure}

\noindent
{{\textbf{Challenge 2:} There have observed significant performance degradation when deploying the trained DONN model to the practical hardware, namely, there is an algorithm-hardware miscorrelation gap between the numerical modeling and the physical system.} The miscorrelation gap can come from two aspects: \textbf{(1)} The imprecise numerical modeling of the DONN system, i.e., the lack of precisely implemented physics emulation intermediate representation (IR). Classic numerical models for fundamental physics kernels in DONNs such as \textit{Finite-difference time-domain} (FDTD) and \textit{scalar diffraction} modeling via \textit{Fast Fourier Transform} (FFT), are both verified to be sufficiently precise in the DONN system emulation \cite{massey2015comprehensive}; 
\textbf{(2)} Lack of domain-specific hardware-software codesign algorithms to realize quantization-aware hardware deployment and deal with the intrinsic noise {(such as fabrication variations, non-unify optical response, etc.)} in optical devices. These challenges have been confirmed by Zhou et al.\cite{zhou2021large} in Figure \ref{fig:intro_correlation}, who reports $\geq 30\%$ accuracy degradation while deploying the trained model to the physical optical system.}

\noindent
\textbf{Challenge 3:} Training and emulation of DONN system are challenging due to high computational cost in modeling the optical physics. For example, \cite{lin2018all,zhou2021large,mengu2022diffractive} reported that training 5-layer DONNs for MNIST-10 with 5 epochs takes 3-4 days (Figure \ref{fig:intro_correlation}). Besides, existing optics simulation frameworks lack runtime optimization in developing the physics kernels, nor domain-specific language (DSL) supports. Table \ref{table:compiler_comp} summarizes the limitations of existing frameworks for DONNs design.
{ More importantly, the choice of numerical physics modeling has significant impacts in runtime efficiency, while it is required to offer high fidelity to the hardware deployment and fabrication.  
}

 
\begin{table}[h]
  \caption{Overview comparisons of existing programming frameworks for DONNs compilation. Lines of Code (LoC) efforts are evaluated with a 5-layer DONNs \cite{lin2018all}.}
  \label{table:compiler_comp}
  \centering
  \begin{tabular}{c|c|c|c|c|c}
    \toprule
        &  \begin{tabular}[c]{@{}c@{}}Optics \\ kernels\end{tabular} &  \begin{tabular}[c]{@{}c@{}}DSE\end{tabular}& \begin{tabular}[c]{@{}c@{}}LoC \\ (val)\end{tabular} & \begin{tabular}[c]{@{}c@{}}LoC \\ (train)\end{tabular} & \begin{tabular}[c]{@{}c@{}}Runtime \\ (pre-fab) \end{tabular} \\
         \midrule
         LightRidge & \cmark & \cmark & 1$\times$ & 1$\times$ & mins -- hrs \\
         \midrule         
         LightPipes\cite{vdovin1997lightpipes} & \cmark &  \xmark & 2$\times$ & n/a & days \\
         \midrule
         \begin{tabular}[c]{@{}c@{}}Customized\\ PyTorch/TF \\ \cite{lin2018all,zhou2021large,mengu2022diffractive}\end{tabular} & \xmark & \xmark & 20$\times$ & 50$\times$ & days\\
    \bottomrule
  \end{tabular}
\end{table}

Thus, we propose \textit{LightRidge}, an agile end-to-end framework, aiming to lower the barriers to design, training, design space exploration, and hardware deployment of DONN systems. 
In particular, \textit{LightRidge} is implemented with high-performance, precise, and versatile optical physics kernels, which precisely correlate to experimental physical systems, enabling out-of-box software-to-hardware realization in an end-to-end fashion, and showing its capabilities to explore advanced DONN architectures for complex ML applications. 
The contributions of this paper are summarized as follows:

\begin{itemize}
    \item We propose a novel agile physics-aware design framework \textit{LightRidge} for end-to-end design, exploration, and deployment for DONNs, consisting of versatile and optimized physics modeling kernels and hardware-software codesign algorithms that enable efficient and precise DONNs modeling w.r.t real-world hardware systems (Section \ref{sec:compiler}). 

    \item We propose LightRidge-DSE to accelerate the end-to-end design cycle for DONNs design, exploration, and on-chip integration, verified by our physical prototype and on-chip integration case study. Moreover, LightRidge-DSE confirms critical domain-knowledge insights \cite{chen2021diffractive} for designing an efficient DONN system in physics meanings (Section \ref{sec:archtecture_variation}). 
    
    

    \item We experimentally validate the effectiveness and precision of LightRidge in designing practical DONN systems and on-chip integration, via visible-range DONN prototype and end-to-end on-chip integration case study (Section \ref{sec:validation}--\ref{sec:integration}).
    
    \item Furthermore, two novel advanced DONN architecture principles are developed via LightRidge to advance DONNs in complex image classification tasks, and first-ever all-optical image segmentation (Section \ref{sec:advance_applications}). 
    
    \item Finally, LightRidge 
    will be released as an open-source hardware project.\footnote{\small \url{https://lightridge.github.io/lightridge}.}
\end{itemize}
 
\begin{figure}
    \centering
    \includegraphics[width=1 \linewidth]{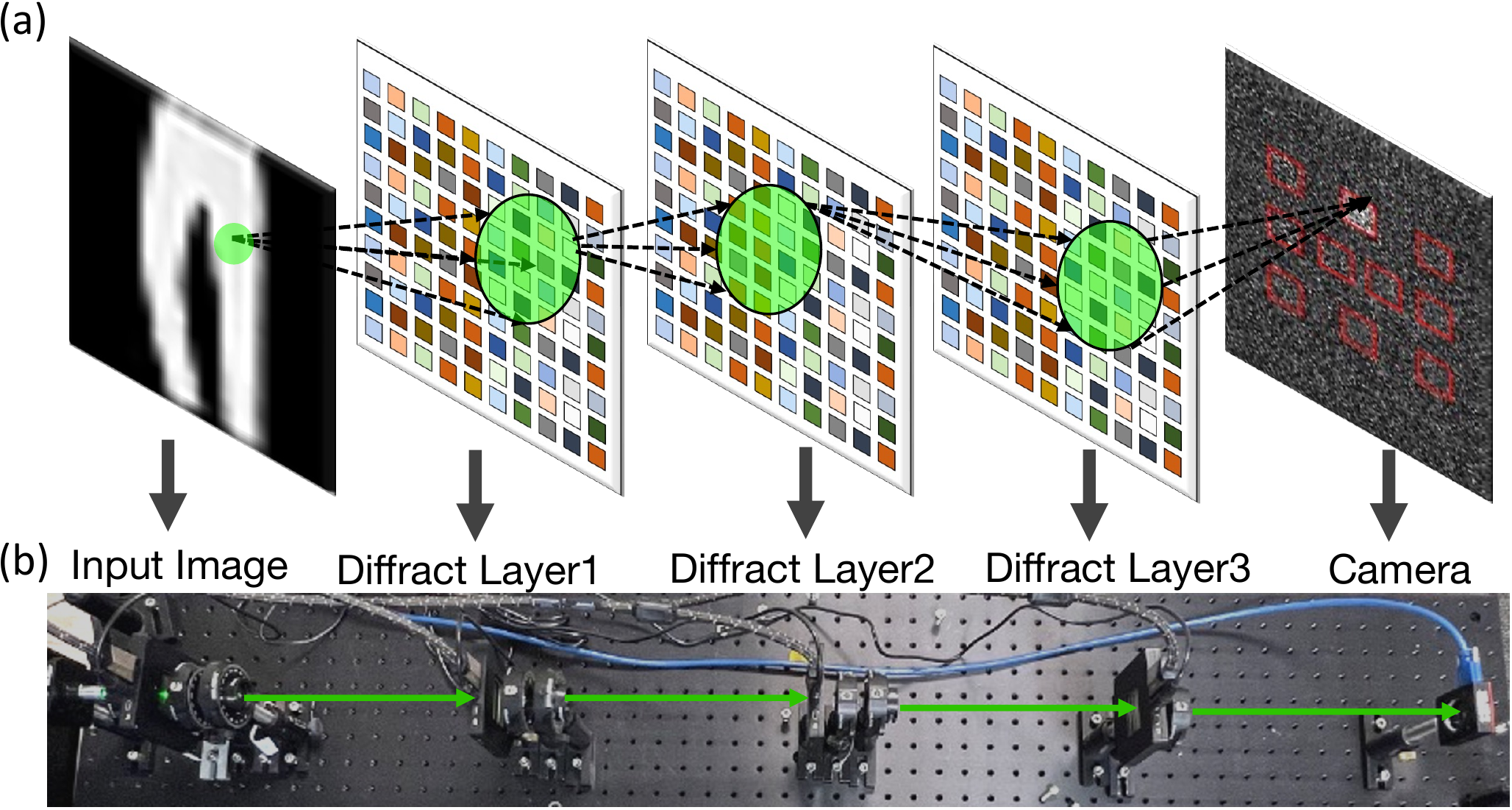}
    \caption{Overview of DONN system and hardware implementation -- \textbf{(a)} Illustration of the DONN system, including the input plane, three diffractive layers, and a light intensity readout plane. \textbf{(b)} The reconfigurable optical hardware system to deploy the DONN system.}
    \label{fig:onn_full}
\end{figure}

%% file: isca2022-latex-template/03sec-donn_2.tex
\section{Diffractive Optical Neural Networks}\label{sec:donn}

{Compared to conventional neural networks (NNs) on digital platforms, the information carrier changes from electrons to photons in DONN systems, i.e., instead of manipulating electrons between transistors to realize the computation, in DONN systems, the computation is realized by manipulating the information-carried light with its physical features. Specifically, the DONN system is composed by multiple diffractive layers stacking in sequence as shown in Figure \ref{fig:onn_full}(a), which embed the phase modulations trained w.r.t the ML task for manipulating and encoding information on the light signal. The connection between layers is realized by the light diffraction when the light signal propagates between layers. {Thus, in DONN systems, light diffraction can be considered as "neural operators" for data movements, and phase change patterns can be seen as "weights" for data manipulations, when compared with conventional NNs.} However, the DONN system requires the analog-to-digital converter to read out the prediction results, where {a detector} is employed at the end of the system to capture the light intensity pattern for analysis and predictions. Thus, DONN systems take advantage of the light signal to encode and propagate information, and its physical nature to realize the computation. Since the physical phenomenon happens by nature with light propagation, the computation happens with no extra energy cost at the light speed for all-optical inference. However, the practical computation efficiency of the DONN system is determined by the analog-to-digital conversion. }

This section presents the overview of DONNs, including emulation, training, and the hardware deployment of DONN systems.
{First, to get an effective DONN model w.r.t a specific ML task, the propagation process of the light signal is emulated and the model is trained based on the optical emulations on digital platforms, where a precise mathematical approximation for the optical phenomenon, i.e., light diffraction and phase modulation, is required, which is illustrated in detail in Section \ref{sec:emulation_kernel}.}
Each point at a given diffractive layer acts as a secondary source of the input light wave in accordance with the Huygens-Fresnel principle. The phase of the input wave is determined by the product of the input wave and the complex-valued phase modulation at that point. The diffraction space is required to generate the diffraction pattern at the receive plane. 
The phase modulation at each point w.r.t its location at the layer is the learnable parameter iteratively adjusted during the training process with error back-propagation method \cite{lin2018all, zhou2021large}. The physical kernel implemented in LightRidge for DONN emulation and training is constructed with the widely used, precise and efficient mathematical approximations for scalar diffraction formulas.
Finally, the trained model is physically deployed with optical devices as shown in Figure \ref{fig:onn_full}(b) or on-chip integration systems as shown in Figure \ref{fig:integration}, to realize the fully optical inference with low energy cost, high computation speed and high system throughput.

\subsection{DONN Emulation and Training}
Enabling the precise hardware-software codesign aware emulation of the physical phenomenon happening in DONN systems including input encoding, light diffraction, phase modulation, and detector reading, is critical for the practical realization of DONN systems. 
There are mainly two mathematical methods for formulating light diffraction: {(1)} \textit{Finite-difference time-domain} (FDTD) method \cite{yee1966numerical}, which performs the full-vector differentiable numerical simulation of photonic structures by solving Maxwell’s equations directly without physical approximations. It is a sophisticated and powerful method for light propagation emulation, while suffering from heavy computation efforts and heavy data dependency that prevent parallelisms in kernel developments. Specifically, FDTD requires the entire computational domain to be sufficiently fine gridded, which means the DONN system size will be expanded exponentially in the FDTD-based emulation. Since DONN systems target large-scale machine learning tasks, the FDTD-based emulation is infeasible in computation runtime and memory for DONN systems due to the system scalability.
{(2)} \textit{Fast Fourier Transform} (FFT) method \cite{goodman2005introduction}, which performs mathematical approximation based on scalar diffraction theory. It simplifies the computation with scenario-specific approximations while keeping the emulation sufficiently precise. There are three widely used approximations for light diffraction in different application scenarios, i.e., \textit{Rayleigh-Sommerfield} approximation, \textit{Fresnel} approximation, and \textit{Fraunhofer} approximation. While both FTDT and FFT-based approximations are differentiable, FFT-based scalar diffraction modeling is more capable for large-scale DONNs emulation without size expansion requirements for fine gridded computational domain. 
More importantly, \cite{massey2015comprehensive} and our physical experiments in Section \ref{sec:evaluation} verify that the FFT-based approximations are sufficiently precise to close the codesign gap for the DONN system emulation. \textbf{Therefore, we implement the FFT-based physics kernel in LightRidge as IR to provide precise and efficient DONN emulation and training (Section \ref{subsec:diff_approx}).} The phase modulation is applied to the input light wave by complex-valued matrix multiplication as illustrated in Section \ref{subsec:phase_mod}. 

In our framework, the FFT based mathematical emulation for light diffraction is design to be fully differentiable from the detector to the laser source w.r.t the loss function acquired from the diffraction pattern captured at the detector. Specifically, during the training process, the prediction is generated according to the intensity of the diffraction pattern captured on the detector with pre-defined detector regions for different classes, where the light intensity $I$ collected by each detector region mimics the probability of output prediction after \texttt{Softmax} in conventional DNNs. Thus, the class whose corresponding detector region collects the highest light intensity is selected as the final prediction. With the {one-hot represented ground truth class} $t$, the loss function $L$ is acquired with the \textbf{MSELoss} between the predictions \texttt{Softmax}~($I$) and one-hot represented ground truth labels $t$, i.e., $L$ = $\parallel$ \texttt{Softmax}($I$) - $t$ $\parallel_{2}$. Thus, the whole system is differentiable and compatible with conventional automatic differential engines.

\begin{figure*}[t]
  \centering
  \includegraphics[width=1 \linewidth]{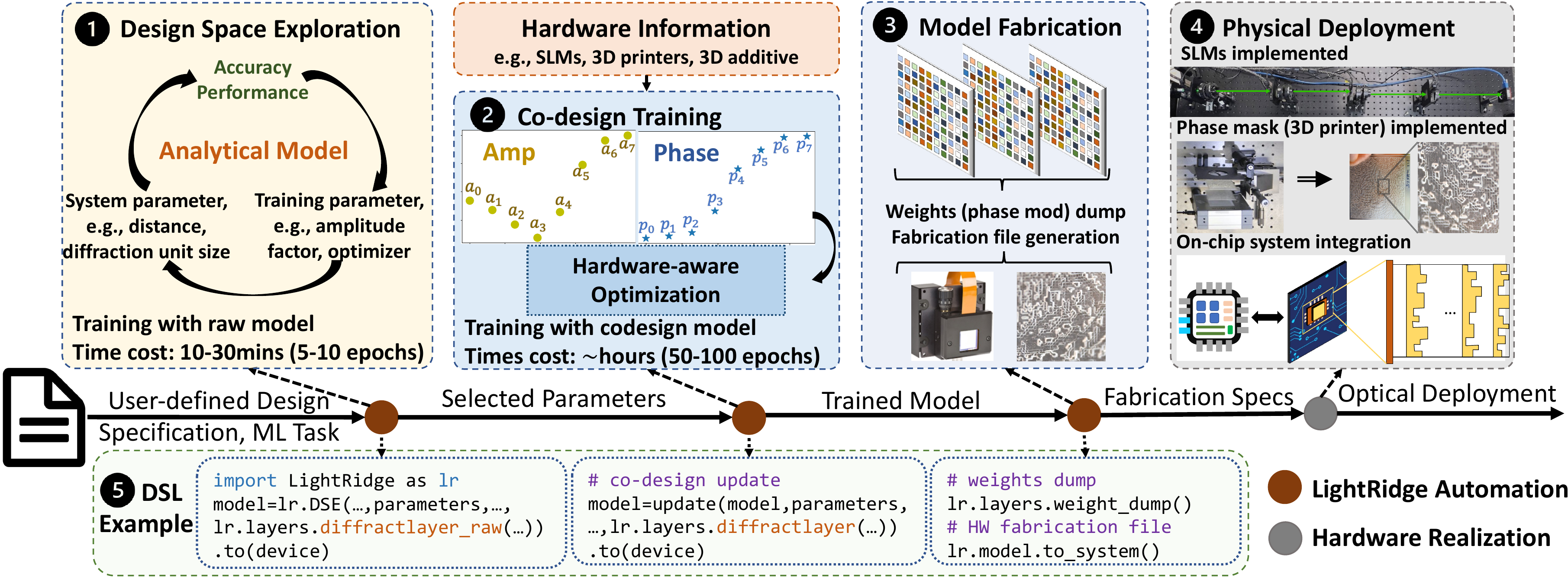}
  \caption{Agile DONN design flow overview -- (1) Design space exploration (DSE) w.r.t the design specification and ML task with LightRidge to automatically search best system parameters; (2) Co-design training with DSE explored parameters and physical hardware/deice parameters; (3) 
  LightRidge backend supports for co-design fabrication; 
  (4) Post-fabrication system integration; (5) LightRidge-DSL that simplifies (1)--(4) with user-friendly front-end APIs. Note that, (1)(2)(3)(5) are executed automatically with LightRidge and (4) is physical demonstrated in this work.}
  \label{figs:overview}
\end{figure*}

\subsection{Hardware Deployment}

The devices for physical hardware to deploy the trained DONN model need to be carefully selected, as optical devices made from different materials can have significantly different optical responses to different laser wavelengths. 
For example, SLMs can function as diffractive layers in the DONN system with the laser wavelength in visible range; while for systems with laser wavelength in Terahertz (THz) range, SLMs cannot provide efficient phase modulations to the light signal and the 3D printed masks with designed thickness at each pixel made with UV-curable resin are used as the diffractive layers in THz optical systems \cite{lin2018all}.

In our experimental hardware systems shown in Figure \ref{fig:onn_full}(b), the wavelength of the laser source is $532$nm, which is in the operating range of the SLM\footnote{ \url{https://holoeye.com/lc-2012-spatial-light-modulator/}}. Specifically, the SLM is an array of twisted nematic liquid crystal, where each pixel (liquid crystal) can be independently twisted to different angles by different applied control voltages, providing different phase modulation for the input light beam. However, such analog optical devices hardly have unified optical response to the control and can vary from each single due to fabrication errors, worsening the correlation between the numerical emulations and the hardware deployment, which highlights the importance to design precise computation kernels for emulation and hardware-software codesign algorithms for DONN systems. 

%% file: isca2022-latex-template/04sec-compiler.tex
\section{LightRidge Framework}\label{sec:compiler}

{Figure \ref{figs:overview} shows the end-to-end design flow of DONN systems with automation provided by LightRidge. With the user-defined design specification and the targeted ML task, \circled{1} the architectural and fabrication parameters such as diffraction distance, diffraction unit size, chip dimensions, etc., are selected and produced automatically by conducting fast and efficient design space exploration (DSE) with the emulation model in LightRidge, which circumvents the critical domain knowledge requirements for designing a functioning DONN model (Section \ref{sec:archtecture_variation}). This exploration is enabled with our accelerated and precise emulation engine, improving the runtime efficiency significantly. \circled{2} When the satisfying hyperparameters are acquired from the fast DSE, the emulation model will be updated with the hardware information for physical deployment, e.g., the optical response curve for SLMs w.r.t the control voltages, where the emulation model is further trained with codesign algorithms with hardware-aware optimizations. \circled{3} Optical devices for practical deployment are fabricated/set w.r.t the parameters in the trained model, i.e., the phase modulations in diffractive layers. The device fabrication information is dumped and generated automatically by LightRidge. \circled{4} With all components ready for deployment, a targeted all-optical DONN system can be setup for efficient and energy-saving all-optical inference (Section \ref{sec:evaluation}). \circled{5} Moreover, the LightRidge automation processes are all efficiently realized by the user-friendly DSL support in LightRidge.

In this section, we will introduce the LightRidge framework including the physics kernel with mathematical approximation modelling for DONN systems implemented in LightRidge, a novel complex-valued regularization algorithm to improve the training performance, and the front-end DSL designed for the LightRidge compilation implementations. 
}

\subsection{Physic Kernel Implementation}\label{sec:emulation_kernel}

The DONN system functions as a neural network based on two physical phenomena, i.e., light diffraction and phase modulation. In our framework, we take FFT-based scalar diffraction theory to build our modelling kernels. 

First, the continuous-wave (CW) laser source is deployed to encode the input information. The light wave is described with complex-valued numbers in physics with two properties, amplitude and phase of the wave, i.e., $E$=$Ae^{j\theta}$, where $j$=$\sqrt{-1}$, $A$ is the amplitude, and $\theta$ is the phase. The input information is encoded with the intensity $I$ of the light wave with phase initialized as $0$, i.e, $\theta\text{=}0$, $A\text{=}I$. Then, as shown in Figure \ref{figs:diffract_ill}, the information-carried light wave is diffracted over the diffraction distance $z$, emulated with mathematical diffraction approximations described in Section \ref{subsec:diff_approx}. At the diffractive layer, each diffraction unit embeds a phase modulator, where the trainable parameter, phase modulation, is applied to the light signal as described in Section \ref{subsec:phase_mod}. 
The forward function for a multiple-layer constructed DONN system calculates diffraction and phase modulation iteratively through all stacked diffractive layers.   
Finally, the diffraction pattern, i.e., the distribution of light intensity, is captured and converted to digital processable information at the detector for computer processing.

\subsubsection{\textbf{Light Diffraction approximation}}
\label{subsec:diff_approx}
 
There are typically three mathematical approximation methods for scalar theory of diffraction, i.e., \textit{Rayleigh-Sommerfeld approximation}, \textit{Fresnel approximation}, and \textit{Fraunhofer approximation}. They work under specific application scenarios with different assumptions of the system, such as aperture size and propagation distance.

The Rayleigh-Sommerfeld is the most commonly used approximation as it works with least physical approximations of the system and is reported to give quite accurate results. The Rayleigh-Sommerfeld approximation is implemented with Equation \ref{equ:sommerfeld} in our framework.
\begin{figure}[]
  \centering
  \includegraphics[width=1\linewidth]{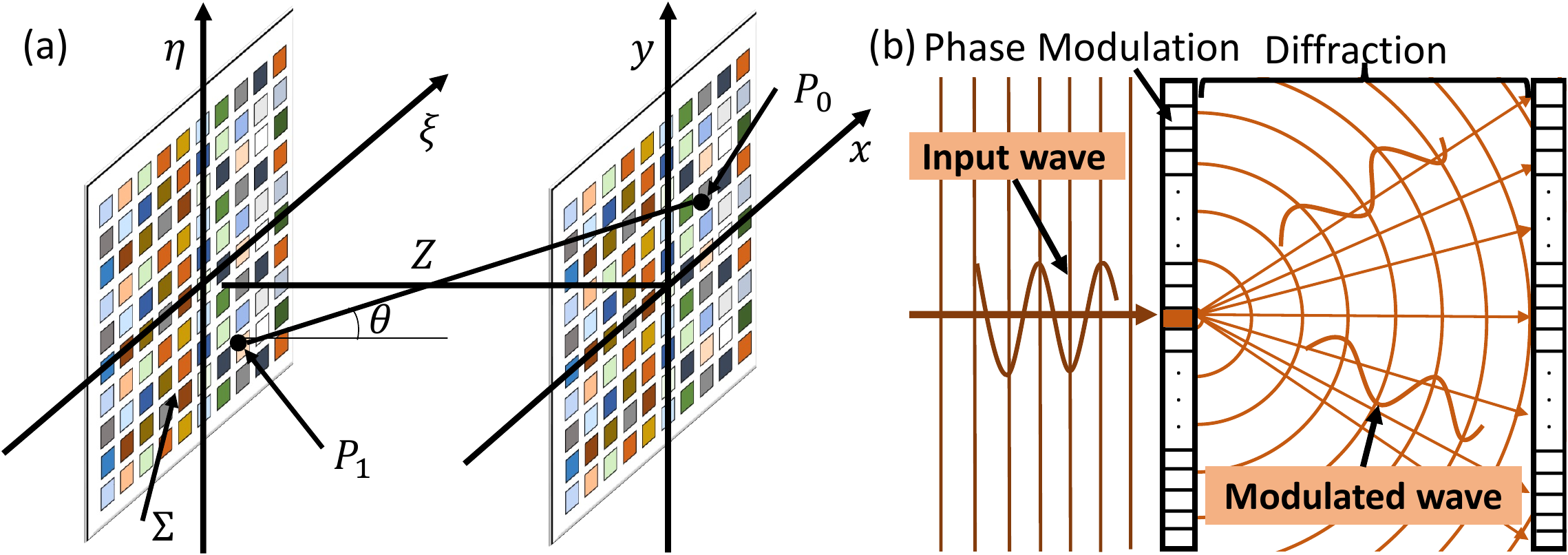}
  \caption{Diffraction illustration. -- \textbf{(a)} ($\xi$, $\eta$) is the plane for diffraction aperture and illuminated by the input light beam in positive $z$ direction, $\Sigma$ on the plane ($\xi, \eta$) denotes the illuminated area. ($x, y$) plane is the target plane. $P_{1}$ and $P_{0}$ are illuminated points on the planes. $\theta$ is the angle between the outward normal and the vector pointing from $P_{1}$ to $P_{0}$. \textbf{(b)} Light propagation and phase modulation through the diffractive layer w.r.t the input light wave.}
  \label{figs:diffract_ill}
\end{figure}
As shown in Figure \ref{figs:diffract_ill}(a), the diffracting aperture is in the ($\xi, \eta$) plane, and is illuminated in the positive $z$ direction. We calculate the wavefield across the ($x, y$) plane, which is parallel to the ($\xi, \eta$) plane and at distance $z$ from it. The $z$ axis pierces both planes at their origins. Then, when $r_{01} \gg \lambda$, the \textit{Rayleigh-Sommerfeld approximation} will be described as
\begin{equation}
\small
    U\text{(}x, y, z\text{)} ~\text{=}~ \frac{z}{j\lambda}\iint \limits_{\Sigma}U\text{(}\xi, \eta, 0\text{)}\frac{\text{exp}\text{(}jkr_{01}\text{)}}{r_{01}^{2}} d{\xi}d{\eta}
\label{equ:sommerfeld}
\end{equation}
where $j$ = $\sqrt{-1}$, $U$($x, y, z$) describes the wavefield on target ($x, y$) plane after diffraction distance $z$ and $U$($\xi, \eta, 0$) describes the wavefield on the emission ($\xi, \eta$) plane, $\lambda$ is the wavelength of the input laser, $k$ is the wave number where $k$=$\frac{2\pi}{\lambda}$, $r_{01}$ is the vector pointing from $P_{1}$ to $P_{0}$ and the distance $r_{01}$ is given by 
\begin{equation}
    r_{01} \text{=} \sqrt{z^{2} + \text{(}x - \xi\text{)}^{2} + \text{(}y - \eta\text{)}^{2}}
\label{equ:r01}
\end{equation}

When diffraction angle $\theta$ shown in Figure \ref{figs:diffract_ill} is small enough, the computation complexity can be further reduced by applying conditions to the application scenarios while maintaining the emulation accuracy. As a result, In \textit{Fresnel approximation}, by simplifying $r_{01}$ with binomial expansion of the square root in Equation \ref{equ:r01} and eliminating terms but $z$ in the $r_{01}^{2}$ appearing in the denominator of Equation \ref{equ:sommerfeld}, it is described as
\begin{equation}
    U(x, y, z) = \frac{e^{jkz}}{j\lambda z} \iint \limits_{\Sigma} U(\xi, \eta, 0)\text{exp}\{j\frac{k}{2z}[(x - \xi)^2 + (y - \eta)^2]\}d\xi d\eta
\label{equ:fresnel}
\end{equation}

In Fresnel approximation, the critical approximation happens in the approximation of the exponent, which can be seen that the spherical secondary wavelets will be replaced by wavelets with parabolic wavefronts. Thus, the condition on the distance $z$ will be $z^{3} \gg \frac{\pi}{4\lambda}[(x - \xi)^{2} + (y - \eta)^{2}]^{2}_{max}$, i.e., the observer (the $(x, y)$ plane) is in the near field of the aperture. 

Furthermore, when $z \gg \frac{k(\xi^{2} + \eta^{2})_{max}}{2}$ is satisfied, which means the quadratic phase factor under the integral sign in Equation \ref{equ:fresnel} is approximately unity over the entire aperture, \textit{Franuhofer approximation} will further greatly simplify the calculations. Thus, in the far field of the aperture, the diffraction can be approximated as 
\begin{equation}
\small
    U(x, y, z) = \frac{e^{jkz}e^{j\frac{k}{2z}(x^{2} + y^{2})}}{j\lambda z}\iint \limits_{\Sigma} U(\xi, \eta, 0) exp[-j\frac{2\pi}{\lambda z}(x\xi + y\eta)]d\xi d\eta
\label{equ:fraunhofer}
\end{equation}

Thus, the diffraction process can be more generally formulated as -- when an input wave resulted from $l-1$-th layer $(\xi, \eta)$, $U_{l-1}(\xi, \eta, 0)$, diffracts over diffraction distance $z$ to the $l$-th layer $(x, y)$, the resulted wavefield $U_{l}^{1}(x, y, z)$ in time domain is described as
\begin{equation}\label{equ:forward_diff}
        U_{l}^{1}(x, y, z) = \iint U_{l-1}(\xi, \eta, 0)h(x-\xi, y-\eta, z)d{\xi}d\eta
\end{equation}
where $h$ is the diffraction function of free space. It can be calculated with spectral algorithm with Fast Fourier Transform (FFT) for fast and differentiable computation. By convolution theorem, the integral can be calculated with
\begin{equation}
    \mathcal{F}_{xy}(U_{l}^{1}(x, y, z)) =  \mathcal{F}_{xy}(U_{l-1}(\xi, \eta, 0))\mathcal{F}_{xy}(h(x, y, z))
\end{equation}
\begin{equation}
\label{equ:diffraction_freq}
    F_{l}(\alpha, \beta, z) = F_{l-1}(\gamma, \sigma, z)H(\alpha, \beta, z)
\end{equation}
Then, the multiplication result $F_{l}(\alpha, \beta, z)$ will be transformed back to the time domain as $U_{l}^{2}(x, y, z)$ by \textit{inverse Fast Fourier Transform} (iFFT) for phase modulation, which is the input wavefunction for applying the phase modulation.

\input{isca2022-latex-template/compiler-table}

\subsubsection{\textbf{Phase modulation}}
\label{subsec:phase_mod}
 
The phase modulation functions like \textit{weight parameters} in conventional neural networks and is updated iteratively during training process. Specifically, the input wave $U_{l}^{2}(x, y)$ (for simplicity, we discard $z$ in phase computation representation as $z$ is not involved) can be described by its amplitude and phase. By \textit{Euler's formula}, it can be described with a complex-valued number in time domain, i.e.,
\begin{equation}
\label{equ:phase_euler}
    U_{l}^{2}(x, y) = A(x, y)e^{j\theta(x, y)} = A\text{cos}\theta + jA\text{sin}\theta
\end{equation}
Where $j=\sqrt{-1}$, $A$ is the amplitude, $\theta$ is the phase of the input wave; Acos$\theta$ is the real part and Asin$\theta$ is the imaginary part. After applying the phase modulation $\phi(x, y)$, the wave function is modulated as:
\begin{equation} \label{equ:phase_mod_matrix}
\begin{split}
    U_{l}(x, y) &= Ae^{j(\theta + \phi)} \\
    &= (A\text{cos}\theta + jA\text{sin}\theta) \times (\text{cos}\phi + j\text{sin}\phi)\\
    &= U_{l}^{2}(x, y) \times \phi(x, y)
\end{split}
\end{equation}
which can be realized with complex-valued matrix multiplications. $U_{l}(x, y)$ is the input wavefunction for the forward function (Equation \ref{equ:forward_diff}) for the $l+1$-th diffractive layer.  

\subsection{Codesign Algorithm with Physics-aware Complex-valued Regularization}
\label{sec:regu}
 
First, for the model emulation and training process on digital platforms, considering the physics in optics, the DONN system is described and emulated with complex-valued numbers. However, according to Equation \ref{equ:phase_mod_matrix}, the training for the DONN system is more phase modulation dominated, while the intensity at the end of diffraction will decrease exponentially as the number of diffractive layers increases, which means a regularization between amplitude and phase is required to avoid gradient vanishing and explosion in the training process. With this insight, we introduce a novel regularization factor $\gamma$ in the forward function to improve the training efficiency, which can flexibly change the gradient scales between amplitude and phase modulations. Specifically, $\gamma$ is applied to amplitude vector $A$ in Equation \ref{equ:phase_mod_matrix}, where $A$ is implemented with $\gamma A$. 

Furthermore, our framework integrates the physics-aware codesign algorithm presented in~\cite{li2022physics} for efficient hardware deployment of the trained DONN model. Specifically, the framework takes the vector of experimentally measured optical responses w.r.t arbitrary optical hardware (e.g., the calibrated optical response of a SLM shown in Figure \ref{figs:overview}\circled{2}) as inputs, which is discrete and can have different levels of available valid optical responses. However, for optical devices, the number of available levels are usually too limited to fit an accurate function curve. With the implemented algorithm, the discrete and level-limited hardware-aware vector is formulated with Gumbel-Softmax~\cite{jang2016categorical} for differentiable training to map the training parameters directly to the available hardware levels during the training process, i.e., quantization-aware training without quantization approximations, which saves the manual calibration efforts and improves the end-to-end DONN deployment efficiency as shown in Figure~\ref{fig:intro_correlation} and Figure \ref{fig:prop_all}.

\subsection{LightRidge Framework} \label{subsec:language}

{LightRidge framework (Table \ref{tab:language}) consists of four major components to simplify and accelerate the process of design, exploration and deployment of the DONN system, including \textbf{a)} versatile programming modules for precise physics modeling, \textbf{b)} domain-specific neural architecture modules of DONNs, \textbf{c)} accelerated physics kernels for training and inference runtime improvements, and \textbf{d)} hardware deployment supports. 

\noindent\textbf{Low-level physics modeling} -- Three components are required to design a DONN model, including laser source, diffractive layers, and optical/photon detector. To model the whole physical phenomenon of DONNs, we first introduce the mathematical modeling modules for the implementation of DONN systems -- \textbf{(1)} Various laser source modelings with flexible wavelength settings and beam profiles.
\textbf{(2)} Precise light diffraction approximation, which falls into three categories -- \textit{Rayleigh-Sommerfeld}, which handles both far and near fields but with the highest computational complexity (Equation \ref{equ:sommerfeld}); \textit{Fresnel}, which approximates the propagation with parabolic wavefronts, namely the near field propagation (Equation \ref{equ:fresnel}); \textit{Fraunhofer}, implemented with Equation \ref{equ:fraunhofer}, approximating the propagation with planar wavefronts in the far-field \cite{tobin1997introduction}. \textbf{(3)} The optical/photon detector digitizes the analog light intensity to make it processable by the computer.  

\noindent\textbf{Model-level APIs} -- The DONN model is constructed with flexible model-level modules with LightRidge, where the architectural parameters can be used to customize the system -- \textbf{(1)} the laser source module \texttt{lr.laser} offers precise laser customization including laser specifications such as \texttt{wavelength}, \texttt{src\_profile}, etc. 
\textbf{(2)} The physics modeling of diffraction with trainable phase modulation is implemented in \texttt{lr.layers}. Two diffraction modelling with and without hardware-aware optimization are provided with \texttt{lr.layers.diffractlayer} and \texttt{lr.layers.diffractlayer\_raw}, repectively. {Specifically, to deal with \textbf{challenge 2} in Section \ref{sec:intro}, \texttt{lr.layers.diffractlayer} employs the codesign algorithm, where the device-level information is delicately integrated in the training process with quantization methods in ~\cite{li2022physics}
applied on the trainable parameters in diffractive layers for efficient modeling-to-hardware deployment. 
} Both modules can alternate three diffraction approximation algorithms according to the user definition. Additionally, user-defined system hyperparameters such as size of diffraction unit (\texttt{pixel\_size}), diffraction distance (\texttt{distance}), the available levels of the hardware implementing diffractive layers (\texttt{level}) can also be customized easily with our framework. 
\textbf{(3)} The detector is employed to capture the light intensity after propagation and modulation through the system, which is the interface component for linking training loss construction and the DONN model emulation. In \texttt{lr.layers.detector}, \texttt{x\_loc} and \texttt{y\_loc} are lists of spatial coordinates of the detector, and the size of the detector regions is customized by \texttt{det\_size}. 
\textbf{(4)} Finally, \texttt{lr.models} is a sequential container that stacks arbitrary numbers of customized diffractive layers in the order of light propagation in the DONN system and a detector plane. As a result, we construct a complete DONN system just like constructing a conventional neural network.


\noindent\textbf{Training support} -- 
The DONN model is trained with conventional automatic differentiation engines in complex domain, which is supported by our differentiable physics kernels and training utility functions. 
Specifically, the original one-dimensional input is processed to a complex-valued input by initializing the phase information in \texttt{data\_to\_cplex}. Training parameters such as \texttt{optimizer}, complex-valued regularization \texttt{regu\_factor}, loss function \texttt{loss}, etc., are also enabled in complex domain by \texttt{lr.train.utils}. The CPU and GPU accelerations are enabled by \texttt{to(device)}. Finally, \texttt{lr.train.dse} enables physics-aware DSE for DONNs design and integration (Section \ref{sec:archtecture_variation}).

\noindent\textbf{Hardware deployment} -- The visualization of trained model parameters is provided with \texttt{lr.layers.view()}. To practically deploy the digitally trained model to hardware, the quantization to the specific hardware (post-training quantization) is provided by \texttt{lr.model.to\_system}. For example, for SLMs implemented DONN systems, the framework produces the trained applied control voltage array for each SLM for light signal manipulations. For THz systems, which is implemented with 3D printed phase masks, the framework will produce the thickness array for mask fabrications by calling \texttt{lr.model.to\_system}.
}

\begin{figure}[t]
    \centering
    \includegraphics[width=0.5\textwidth]{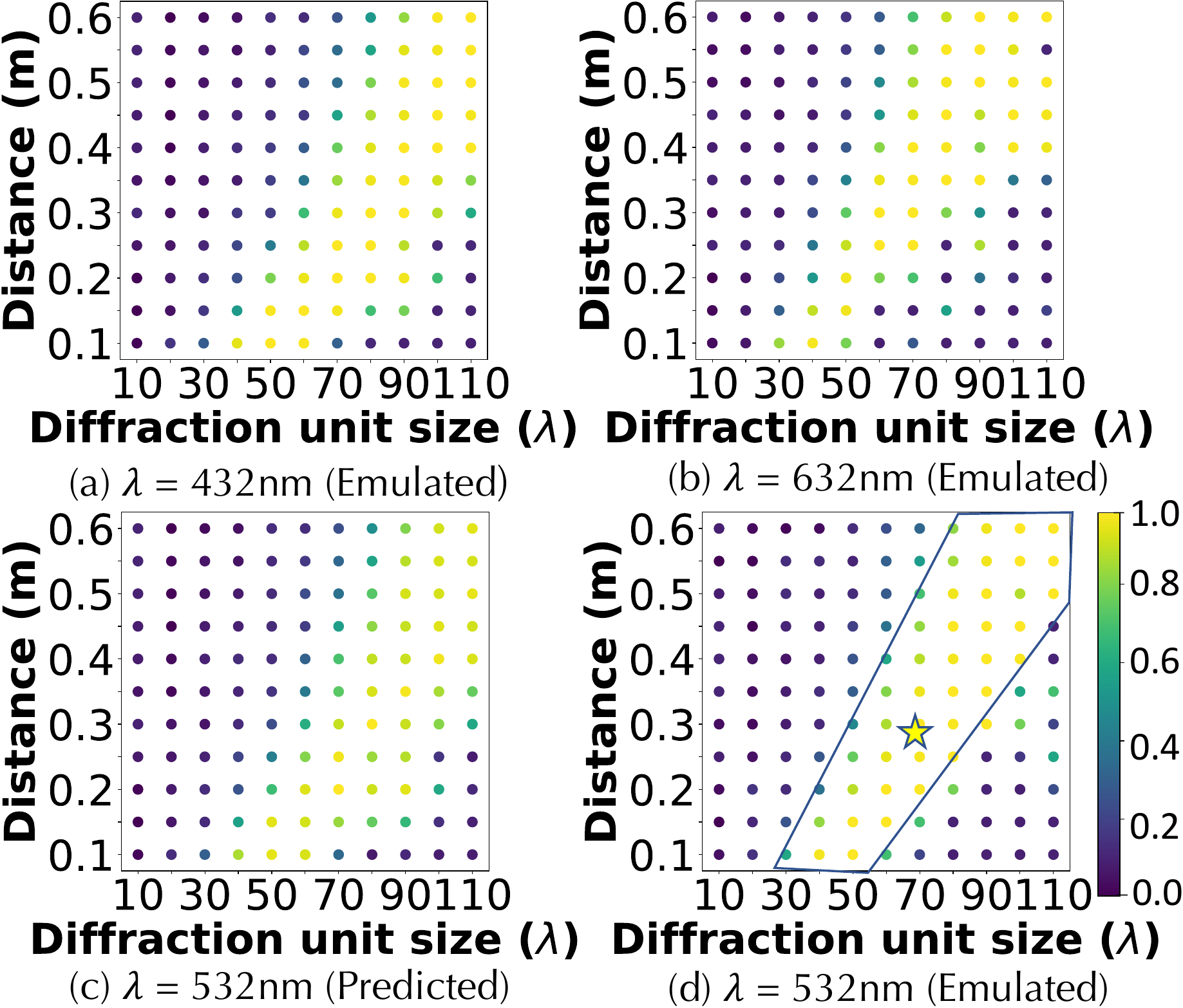}
    \caption{{Results of architectural DSE of DONN systems w.r.t diffractive unit size, and diffraction distance under different laser wavelength ($\lambda$) with each grid colored according to accuracy on MNIST10. \textbf{(a)} and \textbf{(b)} are training data from emulations w.r.t design space under $\lambda=432$nm and $\lambda=632$nm for the inference model.
    \textbf{(c)} Predicted performance w.r.t design space under $\lambda=532$nm with the ML DSE model trained with data points from (a) and (b). \textbf{(d)} Grid-search validation under $\lambda=532$nm that verify the ML-based DSE quality. The DSE-guided setup at the star point is verified with the experimental prototype in Section \ref{sec:evaluation}.}}
    \label{fig:DSE_pixel_dist}
\end{figure}

%% file: isca2022-latex-template/compiler-table.tex
\begin{table*}[h]
\small
    \centering
      \caption{Overview of the LightRidge programming modules and partial front-end APIs. Note that we use \texttt{lr} to represent our integrated Python package \texttt{lightridge}.}
    \label{tab:language}
\begin{tabular}{|p{5em}|p{11em}|p{42em}|}

     \midrule
     Classes & Modular Programs & Description\\
     \midrule
     
     \multirow{3}{*}{\begin{tabular}[c]{@{}c@{}}\\ \\ \\ ~~Low-level \\ ~~~modeling\end{tabular}} & Laser source \& profiles & Modeling coherent laser beams with various wavelength/profiles, e.g., Gaussian beam, Bessel beam, etc. \\
     
     \cmidrule(r){2-3}
     
     & Diffraction approximation & High-performance tensor implementations of numerical diffraction approximations, including \textbf{Rayleigh-Sommerfeld} (Equation \ref{equ:sommerfeld}), \textbf{Fresnel} (Equation \ref{equ:fresnel}), and \textbf{Fraunhofer} (Equation \ref{equ:fraunhofer}).  \\
     \cmidrule(r){2-3}
     
     & Optical/photon detector & An photon detector to capture the light intensity and convert the analog intensity information to the digital computer-processable information. \\
     \midrule
 
     \multirow{4}{*}{ \begin{tabular}[c]{@{}c@{}} \\ \\ \\ Model-level\\ APIs\end{tabular}} & \texttt{lr.laser} & Define the laser source for the system, including laser wavelenghth and its beam profile. \\
     \cmidrule(r){2-3}
    & \vspace{1mm}\texttt{lr.layers} & Include modules of different types of diffraction modeling, e.g., hardware-specific layer module \texttt{lr.layers.diffractlayer} and general diffractive layer \texttt{lr.layers.diffractlayer\_raw}, that can be configured with various approximation methods, distance, diffraction unit size, etc. \\
    \cmidrule(r){2-3}
    & \texttt{lr.layers.detector} & Define detector designs for various ML tasks, e.g., in image classification task, the coordinate and the size of the detection region for each candidate class. \\
    \cmidrule(r){2-3}
    & \texttt{lr.models} & Sequential container to customize DONN system by stacking diffractive layer and detector modules.\\
     \midrule

     \multirow{4}{*}{\begin{tabular}[c]{@{}c@{}} \\  ~~~Training \end{tabular}} &  \vspace{0.2mm}\texttt{lr.train.utils} & Training utility modules including data handling (e.g., \texttt{utils.data\_to\_cplex}), complex-valued regularization, loss function, optimizer, etc.\\
     \cmidrule(r){2-3}
         & \texttt{lr.train.to(device)} & Enable CPU and GPU accelerations for accelerating diffraction emulation and DONNs training.  \\ \cmidrule(r){2-3}
          & \texttt{lr.train.dse(specs)} & Perform pre-fabrication design space exploration with chip integration specifications as inputs.   \\
        \midrule
    \multirow{2}{*}{\begin{tabular}[|c|]{@{}c@{}}  Hardware \\ deployment\end{tabular}}  & \texttt{lr.layers.view()} & Visualize the original phase value per layer or values w.r.t the hardware specifications. \\
    \cmidrule(r){2-3}
    
    &\texttt{lr.model.to\_system} &  Generate device-specific phase parameters for deployment w.r.t the hardware specifications (e.g., configurations of SLMs, thickness of 3D printed masks). \\
     \midrule
\end{tabular}
\end{table*}

%% file: isca2022-latex-template/05sec-DSE.tex
\section{{Design Space Exploration}}
\label{sec:archtecture_variation}

Taking advantages of LightRidge, we introduce the first explicit architectural design space exploration (DSE) engine for DONNs, namely LightRidge-DSE. As discussed earlier, the domain knowledge of optics and optical hardware are critical technical barriers to design DONNs. Therefore, there is a great need to enable an automatic DSE exploration in LightRidge, which will significantly shorten the design and hardware deployment cycle of DONNs and lower the optical domain-knowledge requirements. We propose an analytical model based DSE approach to accelerate the DSE process, where the analytical model is extracted from a ML regression model. Our main goal of the DSE engine is to provide guidance to design DONN systems under new design parameters with fabrication and chip integration requirements (e.g., fabrication technologies, chip dimension, etc.) with learnt knowledge from existing setups.

{ 
\noindent
\textbf{Design space of DONNs} -- We consider the DONN design space from two aspects: (1) The major physical architectural design parameters of DONNs include -- {\circled{1} the diffraction unit size (the dimension of each diffractive unit), and \circled{2} the diffraction distance, i.e., the physical distance between the source to the first diffractive layer, layer to layer and the last layer to the detector ($z$ in Figure \ref{fig:onn_full}). These two are critical architectural parameters under a fixed laser profile (wavelength). (2) The space exploration over DONNs, i.e., spatial architectural parameters -- \circled{3} system size (or system resolution) paired with \circled{4} hardware/device precision, i.e., discrete phase modulation levels provided by the device,} which are sensitive parameters w.r.t the performance of ML tasks. We take the physical architectural DSE as an example in this section.
}

\noindent
\textbf{DSE features and data collection} -- {In our case, we show the process of conducting the DSE with the physical architectural design parameters, i.e, the diffraction unit size $d$ and the diffraction distance $D$, for DONN systems under different laser wavelength $\lambda$. 
With fixed system size $200 \times 200$ and device precision, $256$ optical states covering [0,2$\pi$] for phase modulation, we collect training data by sweeping diffraction unit size from 10$\lambda$ to 110$\lambda$ and diffraction distance $D$ from $0.1$m to $0.6$m on a 5-layer DONN system, i.e., 121 data points, for laser wavelength $\lambda$ of 632nm and 432nm.}

\noindent
\textbf{Analytical model based DONN DSE} -- We employ a gradient boosting regression \cite{prettenhofer2014gradient} model to find out a polynomial analytical model to bypass and transfer optical physics-aware DONNs DSE knowledge to new nearby $\lambda$. Specifically, our analytical model is trained with diffraction unit size and diffraction distance exploration data points from systems with $\lambda=632$nm and $\lambda=432$nm (Figure \ref{fig:DSE_pixel_dist} (a) and (b)) to estimate the DONNs design space in ML performance given a different laser profile with $\lambda=532$nm (Figure \ref{fig:DSE_pixel_dist}(c)). The regression model takes the wavelength $\lambda$, $d$, and $D$ as inputs, and predicts (regression) the accuracy w.r.t MNIST dataset, trained with mean squared error (MSE) loss. {The regression model is built with \textit{n\_estimators=3500, learning\_rate=0.2, max\_depth=3, random\_state=25}}.
The approximated prediction result from the analytical model is employed to guide DONN DSE under a new target $\lambda$. To evaluate the analytical model based DSE strategy, we compare the predicted design space (Figure \ref{fig:DSE_pixel_dist}(c)) with the emulation verified design space (Figure \ref{fig:DSE_pixel_dist}(d)) under $\lambda=532$nm. The star point in Figure \ref{fig:DSE_pixel_dist}(d) shows our analytical DSE can find the best design points, which is further verified by the end-to-end LightRidge development process in Section \ref{sec:evaluation}.


{The analytical model by DSE can generalize the learnt optical formula to DONN systems with new laser wavelength while following the traditional maximum half-cone diffraction angle theory \cite{chen2021diffractive}, i.e., the analytical model should be applied to a nearby wavelength within the applicable range by the theory of the training data. In our DSE example (Figure \ref{fig:DSE_pixel_dist}), we use the analytical model trained from 432 nm and 632 nm for predictions under 532 nm. However, such a analytical model trained with wavelength in visible range will not work for predictions for wavelength in other ranges, such as Infrared (IR) and Microwaves because of the theory violation.}

{
\noindent\textbf{Sensitivity analysis} -- We perform single parameter control variable tests for all three parameters in Table \ref{tbl:dse_sensitivity}. Our results show that diffraction unit size is the most sensitive parameter, while wavelength and distance are almost equally sensitive to the accuracy performance w.r.t the image classification task with MNIST dataset. By shifting the DSE explored best parameters (the star point in Figure \ref{fig:DSE_pixel_dist}(d)) +10\%/+5\% or -10\%/-5\%, we observe sharply accuracy drops on unit size shifting (dropped to 30\% in accuracy by shifting only $\pm$5\%), while less accuracy drops on the other two parameters (dropped to $\sim$70\% in accuracy by shifting $\pm$5\%).

\begin{table}

\small
\begin{tabular}{|c|c|c|c|c|c|}
\hline
{Wavelength} & {\begin{tabular}[c]{@{}c@{}}480 nm \\ (-10\%)\end{tabular}}  & { \begin{tabular}[c]{@{}c@{}}505 nm\\ (-5\%)\end{tabular}}   & {\begin{tabular}[c]{@{}c@{}}532 nm \\ (0\%)\end{tabular}} & {\begin{tabular}[c]{@{}c@{}}560 nm\\ (+5\%)\end{tabular}}    & {\begin{tabular}[c]{@{}c@{}}585 nm\\ (+10\%)\end{tabular}}   \\ \hline
{Accuracy}        & {0.34}& { 0.70}& { 0.97}       & {0.72}& {0.35}\\ \hline
{Distance}       & { \begin{tabular}[c]{@{}c@{}}0.27 m \\ (-10\%)\end{tabular}}  & {\begin{tabular}[c]{@{}c@{}}0.285 m \\ (-5\%)\end{tabular}} & {\begin{tabular}[c]{@{}c@{}}0.30 m \\ (0\%)\end{tabular}} & {\begin{tabular}[c]{@{}c@{}}0.315 nm \\ (+5\%)\end{tabular}} & {\begin{tabular}[c]{@{}c@{}}0.33 nm\\ (+10\%)\end{tabular}}  \\ \hline
{Accuracy}        & {0.33}& {0.70}& {0.97}& {0.74}  & {0.34} \\ \hline
{Unit size} & {\begin{tabular}[c]{@{}c@{}}32.4 um \\ (-10\%)\end{tabular}} & {\begin{tabular}[c]{@{}c@{}}34.2 um\\  (-5\%)\end{tabular}} & {\begin{tabular}[c]{@{}c@{}}36 um \\ (0\%)\end{tabular}}  & {\begin{tabular}[c]{@{}c@{}}37.8 um\\  (+5\%)\end{tabular}}  & {\begin{tabular}[c]{@{}c@{}}39.6 um \\ (+10\%)\end{tabular}} \\ \hline
{Accuracy}        & {0.09}& {0.30}& {0.97}& {0.36} & {0.15}\\ \hline
\end{tabular}
\caption{Sensitivity analysis w.r.t wavelength, diffraction distance, and the diffraction unit size.}
\label{tbl:dse_sensitivity}
\end{table}
}

{With the guidance from the analytical model, LightRidge-DSE finds the best architecture dimension and training parameters with several emulation iterations for selected possible parameters instead of sweeping through the grid-based search space. For example, in our case shown in Figure \ref{fig:DSE_pixel_dist}, aided by the analytical model, few emulation iterations (e.g., two emulations) instead of grid-searching over 121 data points are required for DSE, resulting in 60$\times$ speedups. {On the other hand, DSE engine is able to provide general design parameters for the similar type of ML task. For example, the DSE model for image classification trained by MNIST dataset is also confirmed to be applicable to other MNIST-like datasets such as FashionMNIST \cite{xiao2017fashion}, Kuzushiji-MNIST \cite{clanuwat2018deep}, Extension-MNIST-Letters \cite{cohen2017emnist}~\cite{li2023rubik}.}}

\begin{figure}
     \centering
     \begin{subfigure}[b]{1\linewidth}
         \centering
         \includegraphics[width=\linewidth]{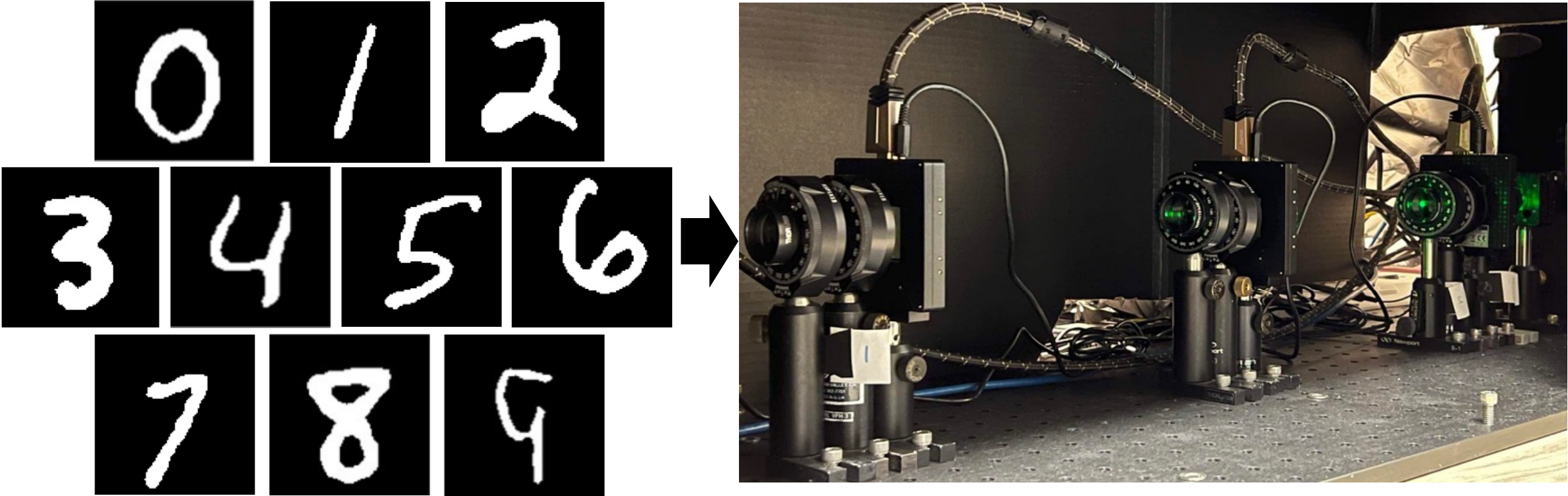}
         \caption{Physical experimental DONNs prototype and measurements.}
         \label{fig:prop_exp}
     \end{subfigure}
     \hfill
     \begin{subfigure}[b]{1\linewidth}
         \centering
         \includegraphics[width=\linewidth]{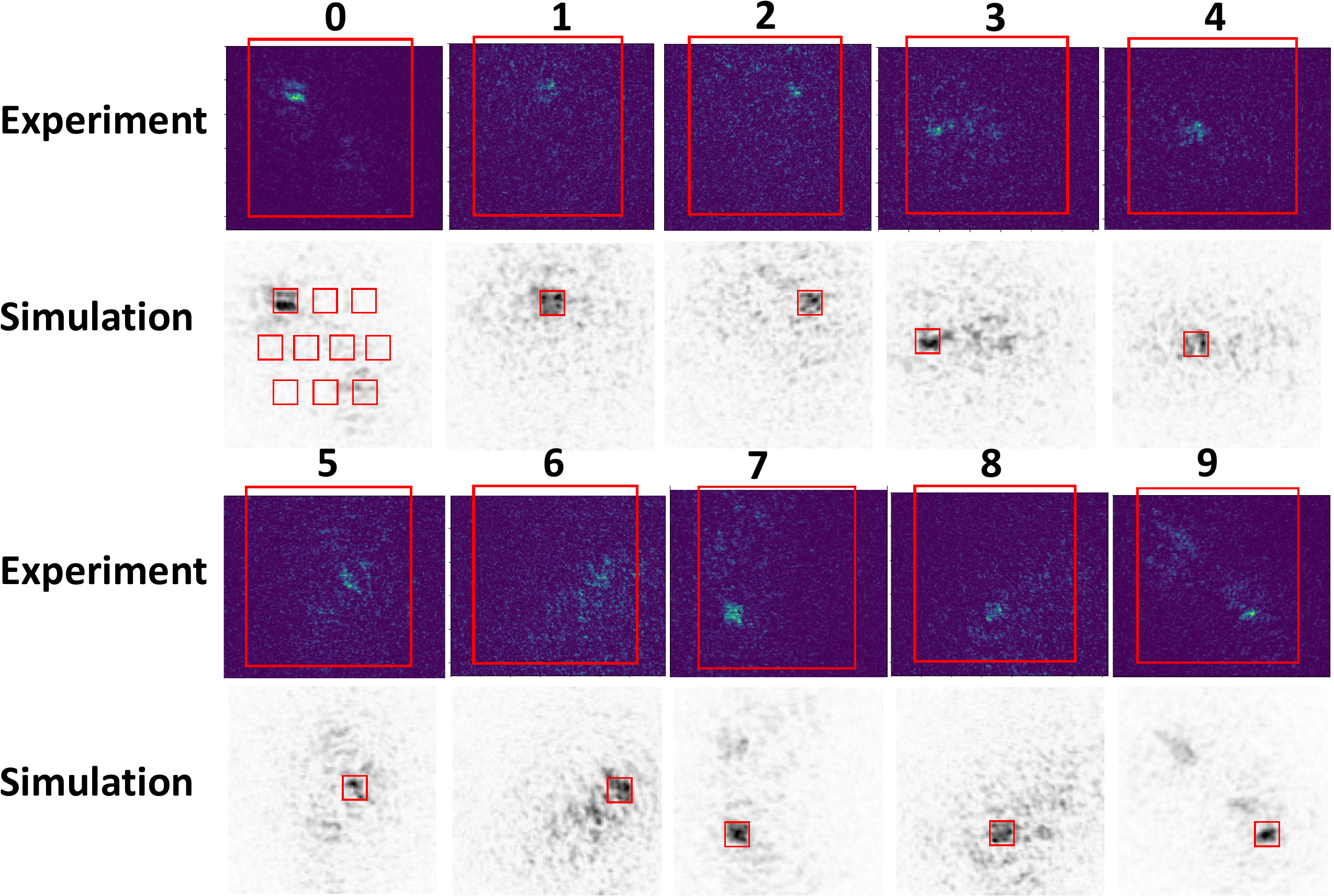}
         \caption{Detector pattern for experiments and simulation.}
         \label{fig:prop_sim}
     \end{subfigure}

        \caption{Evaluation results of a 3-layer DONN system in visible range explored, trained, and deployed by LightRidge -- (a) The experimental system trained and deployed by LightRidge. The corresponding detector pattern from experiments and simulation results produced by LightRidge are shown in Figure \ref{fig:prop_sim}; (b) Corresponding detector patterns of experimental measurements and simulation results (the simulation results generated with \texttt{lr.layers.view()}) of the 3-layer DONN system.
        }
        \label{fig:prop_all}
\end{figure}

%% file: isca2022-latex-template/07sec-results.tex
\section{Evaluation}
\label{sec:evaluation}
\begin{figure}
    \centering
    \includegraphics[width=1\linewidth]{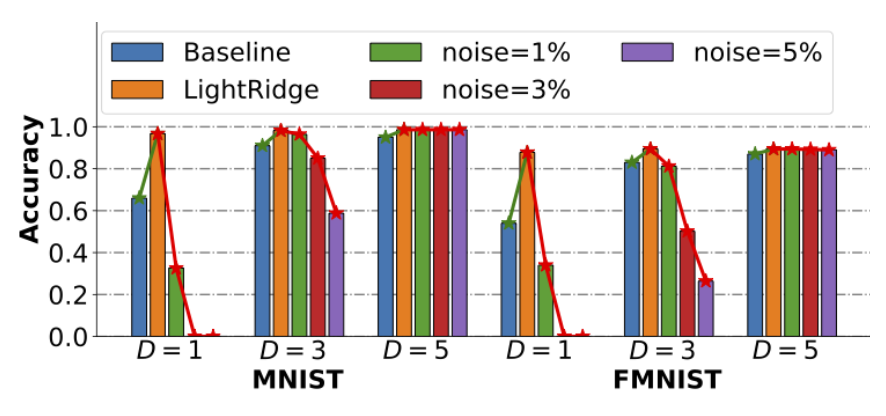}
    \caption{Confidence evaluation of DONNs trained with complex-domain regularization under various system complexity. Baseline results are conducted on methods in \cite{lin2018all,zhou2021large} without noise assumptions.}
    \label{fig:result_acc_conf}
\end{figure}

In this section, we first demonstrate that LightRidge and LightRidge-DSE offer precise hardware-software correlations w.r.t real-world DONNs system realization (Figure \ref{fig:prop_all}) via a visible range DONN prototype. 
Second, we demonstrate the effectiveness of LightRidge framework over SOTA experimental baselines \cite{lin2018all,zhou2021large} in training performance and emulation runtime (Figure \ref{fig:result_acc_conf} -- \ref{fig:runtime_breakdown}).  
Finally, we demonstrate that LightRidge and LightRidge-DSE enables comprehensive DONN system on-chip integration (Figure \ref{fig:integration}) and the capabilities to design advanced DONNs design principles, including multi-channel DONNs classifier on Place365 \cite{zhou2017places} dataset (Figure \ref{fig:rgb_arch}) and the all-optical image segmentation architecture (Figure \ref{fig:seg}). {Note that experiments in Section \ref{sec:validation} are physically deployed on optical hardware shown in Figure \ref{fig:prop_exp}, while other results are from emulations with LightRidge.} 

{


\subsection{LightRidge and LightRidge-DSE Validations via Physical DONNs Prototyping} 
\label{sec:validation}

\noindent
\textbf{Model construction via LightRidge-DSE and training} -- This section demonstrates the hardware-software codesign precision and the effectiveness of LightRidge-DSE, where the parameters of the DONNs model used for physical validation experiments are automatically produced by LightRidge-DSE with the system size of $200 \times 200$. {Specifically, the emulation model for DONN training is constructed with 3 sequentially stacked diffractive layers in \texttt{lr.model}, where each layer is defined with \texttt{lr.layers.diffractlayer} integrating hardware specifications: }
\circled{1} the diffraction pixel size is $36$um $\times$ $36$um; \circled{2} the laser wavelength is $532$nm. Consulted on DSE results shown in Figure \ref{fig:DSE_pixel_dist}(c), distance is explored to be $\sim0.3$m, which is further adjusted to $11$ inches ($0.28$m) on our optical table. There are $10$ pre-defined detector regions for labels placed evenly on the detector plane. The model is trained with MSE loss with one-hot represented ground truth labels using \texttt{Adam} \cite{kingma2014adam} as the training optimizer. The learning rate for the training process is set as $0.5$, the training epoch is set as $100$, and the training batch size is set as $500$ for all experiments.  

\noindent
\textbf{Hardware prototype and validation} -- Laser source CPS532 from Thorlabs, Inc. is implemented as the laser source for the physical DONN system, where SLMs (LC 2012 HOLOEYE) is implemented as diffractive layers. The levels of SLMs for model training are experimentally measured and cover a phase modulation range close to $[0, 2\pi]$. The final diffraction pattern is captured on a CMOS camera (CS165MU1 Thorlabs, Inc.). To make the input easier for hardware deployment, we train and validate the model with {binarized MNIST images} as shown in Figure \ref{fig:prop_exp}, where the trained phase modulation parameters are loaded on the SLMs.


The resulted detector patterns for the inputs are shown in Figure \ref{fig:prop_sim}. The SLM used to encode input binary images is illuminated by the laser source, and the input information will be encoded on the intensity of the input light signal. The intermediate propagation results in all-optical DONN inference are not available as the information is carried with the light beam. At the end of the system, a detector is implemented for analog-to-digital conversion to capture the diffraction pattern, i.e., the light intensity distributions, for model analysis and predictions. As shown in Figures \ref{fig:prop_sim}, DONNs emulation results in LightRidge precisely match the experimental measurements, which demonstrates: (1) precise correlations between the implemented high-level modeling and low-level physics experimental system, which improves the design efficiency significantly without manual HW calibration requirements shown in Figure \ref{fig:intro_correlation}; (2) and the effectiveness of LightRidge-DSE in exploring architecture parameters, which has been further utilized for on-chip integration (Section \ref{sec:integration}).

\subsection{Emulation-level Evaluation}
\label{sec:emulation_acc}

We further verify the design parameters from DSE model as discussed in Section \ref{sec:validation} at emulation level. The accuracy results for image classification with MNIST \cite{lecun1998mnist} and FashionMNIST (FMNIST) \cite{xiao2017fashion} dataset are shown in Figure \ref{fig:result_acc_conf}, where the baseline results are conducted on training methods in \cite{lin2018all}, \cite{zhou2021large} without the proposed physics-aware complex-valued regularization}. The inputs are encoded with the amplitude of the laser beam. To make the input fit the DONN system, we first extend the image with the original size of $28 \times 28$ in MNIST10 and FMNIST datasets to $200 \times 200$ in SLM resolution, and transfer the original one-dimensional image to complex-valued image in the emulation. \textbf{With the regularization factor $\gamma$ implemented, our training algorithm has a significant advantage in training less complex DONN models.} For example, when the DONN model is implemented with only one diffractive layer (D=1), the accuracy performance is $31$\% ($34$\%) improved for MNIST (FMNIST) dataset, compared with the baseline. Additionally, our algorithm can achieve a similar accuracy performance (0.98 for MNIST, 0.89 for FMNIST) for DONN systems regardless of its complexity, i.e., the number of diffractive layers implemented in the system, by adjusting $\gamma$ for the model training.
However, according to the discussion in \cite{lin2018all}, the performance of DONNs with fewer number of layers are fundamentally limited by the optical physics, which is opposite to our accuracy results. 

To understand the increase of accuracy, we analyze the robustness of the DONNs trained with complex-valued regularization. Specifically, we explore the confidence of the predictions acquired by the system, by adding random uniform noise at the detector phase with upper bound 1\%, 3\%, and 5\% intensity noise.
As a result, for both datasets, \textbf{as the depth of DONNs increases, the prediction confidence increases, while the prediction accuracy with no noise applied are all relatively the same.} For example, there is no accuracy degradation on five-layer DONNs for MNIST, and less than 1\% degradation on FMNIST with up to 5\% applied noise. However, for single-layer DONNs, the accuracy drops 63\% for MNIST and 54\% for FMNIST with 1\% noise applied, and drops to 0 when applied noise increases to 3\% and 5\%.


\subsection{LightRidge Runtime Evaluation}

\begin{figure}[!htb]
    \centering
    \includegraphics[width=1.0\linewidth]{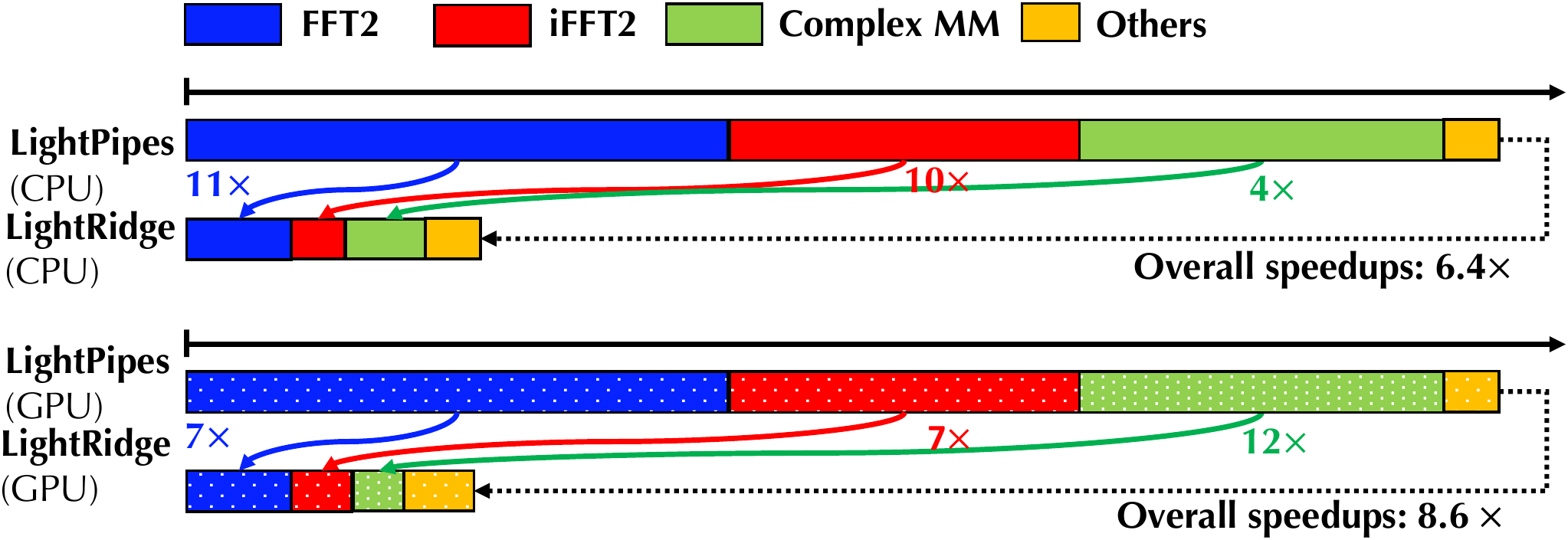}
    \caption{Runtime speedups breakdown with 5-layer 500$\times$500 DONNs. FFT2, iFFT2, and Complex MM are the main operators for DONN numerical modelling.}
\label{fig:runtime_breakdown}
\end{figure}

Runtime efficiency of emulating DONNs is crucial in simulation, training, and exploration. Thus, optimizing runtime performance is another key contribution in LightRidge framework. As shown in Figure \ref{fig:runtime_breakdown}, we first analyze the DONN workloads, where we identify that the majority ($\geq$90\%) of the runtime complexity comes from the numerical modeling of light diffraction. Thus, the major optimization efforts should lie over the diffraction kernels. Second, to effectively utilize the modern computing platforms, we aim to maximize the parallelism from the fundamental physics modeling, which is the main reason of implementing scalar diffraction modeling instead of FDTD in the computation kernel as mentioned earlier in Sections \ref{sec:intro} and \ref{sec:donn}. The diffraction approximation functions with scalar diffraction modeling (Equations \ref{equ:sommerfeld} -- \ref{equ:fraunhofer}) can be breakdown into three major tensor-level operators: complex-domain 2-D FFT (FFT2), inverse 2-D FFT (iFFT2), and complex matrix multiplications (Complex MM). Based on the analysis and kernel breakdowns, we take advantages of modern CPU and GPU platforms by incorporating efficient complex-tensor datatypes and operators. For CPU, the diffraction kernel is optimized via Intel Math Kernel Library (MKL-DNN) complex kernels with AVX-512 support; for GPU, cuFFT, cuFFTW, and cuTENSOR libraries with efficient complex-domain FFTs and MM are deployed. 

To demonstrate the runtime improvements, we compare the runtime of our proposed framework with the commercial tool LightPipes(2021) with its up-to-date version, running various emulation loads, i.e., \{1,3,5,7,10\}-layer DONNs with system resolution sweeping from 100$\times$100 to 500$\times$500. 
All LightPipes-CPU and LightRidge-CPU results are conducted on Intel Xeon Gold 6230 20x CPU. To make fair GPU comparisons, we re-implement the kernels in LightPipes with \texttt{cupy} \cite{cupy_learningsys2017}, and runtime results are collected on Nvidia 3090 Ti GPU platform. 

\begin{figure}[!htb]
     \centering
     \begin{subfigure}[b]{0.49\linewidth}
         \centering
         \includegraphics[width=\linewidth]{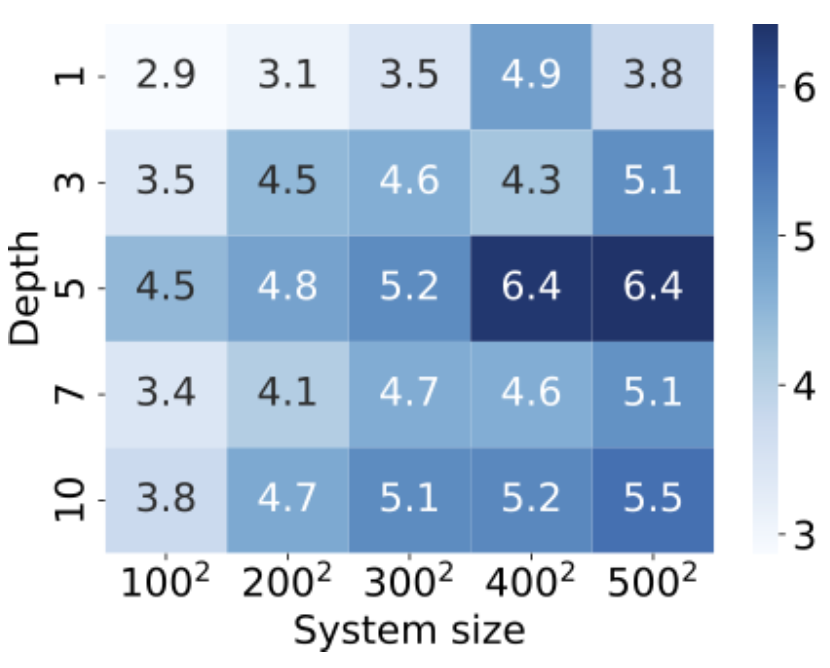}
         \caption{CPU speedups.}
         \label{fig:speedup_CPU}
     \end{subfigure}
     \hfill
     \begin{subfigure}[b]{0.49\linewidth}
         \centering
         \includegraphics[width=\linewidth]{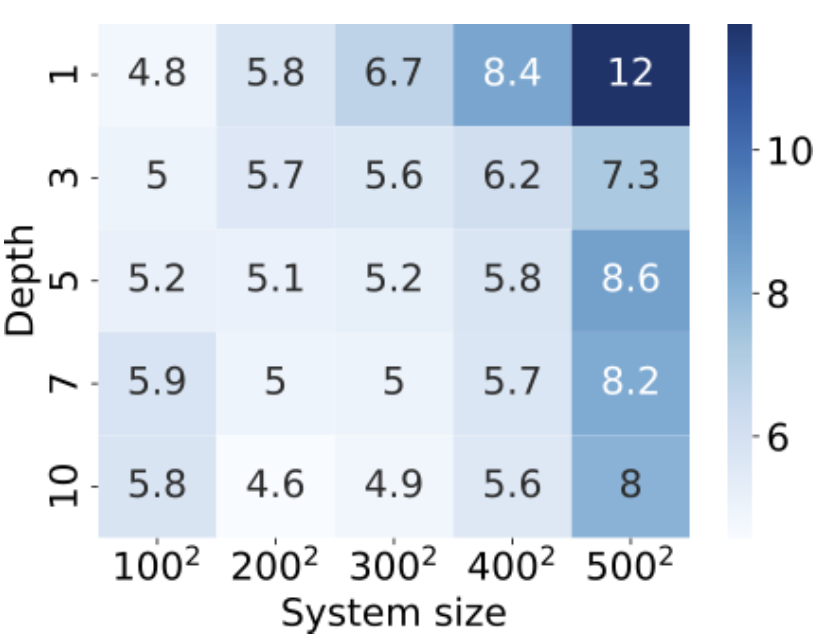}
         \caption{GPU speedups.}
         
         \label{fig:speedup_GPU}
     \end{subfigure}
        \caption{LightRidge runtime speedups over LightPipes with various DONNs system sizes -- \textbf{(a)} CPU speedups. \textbf{(b)} GPU speedups.}
        \label{fig:runtime_speedup}
\end{figure}

Figure \ref{fig:runtime_speedup} shows LightRidge consistently outperforms LightPipes on both CPU and GPU backends. Specifically, Figure \ref{fig:speedup_CPU} shows at most $6.4 \times$ speedup of LightRidge-CPU over LightPipes-CPU at \textit{depth=5, system size=$500^{2}$}. Figure \ref{fig:speedup_GPU} shows at most $12\times$ speedup of LightRidge-GPU over LightPipes-GPU at \textit{depth=1, system size=$500^{2}$}. To understand the runtime speedups offered by LightRidge, we provide normalized speedups breakdown analysis w.r.t LightPipes CPU/GPU, shown in Figure \ref{fig:runtime_breakdown} with 5-layer DONNs workload. We observe that the 6.4$\times$ CPU runtime speedups are contributed from the FFT2 (11$\times$), iFFT2 (10$\times$), and Complex MM kernels (4$\times$); similarly for GPU, {8.6$\times$} overall speedups are primarily contributed from the three kernels with 7$\times$, 7$\times$, and 12$\times$ speedups, respectively. 

{
\begin{figure}
    \centering
    \includegraphics[width=1\linewidth]{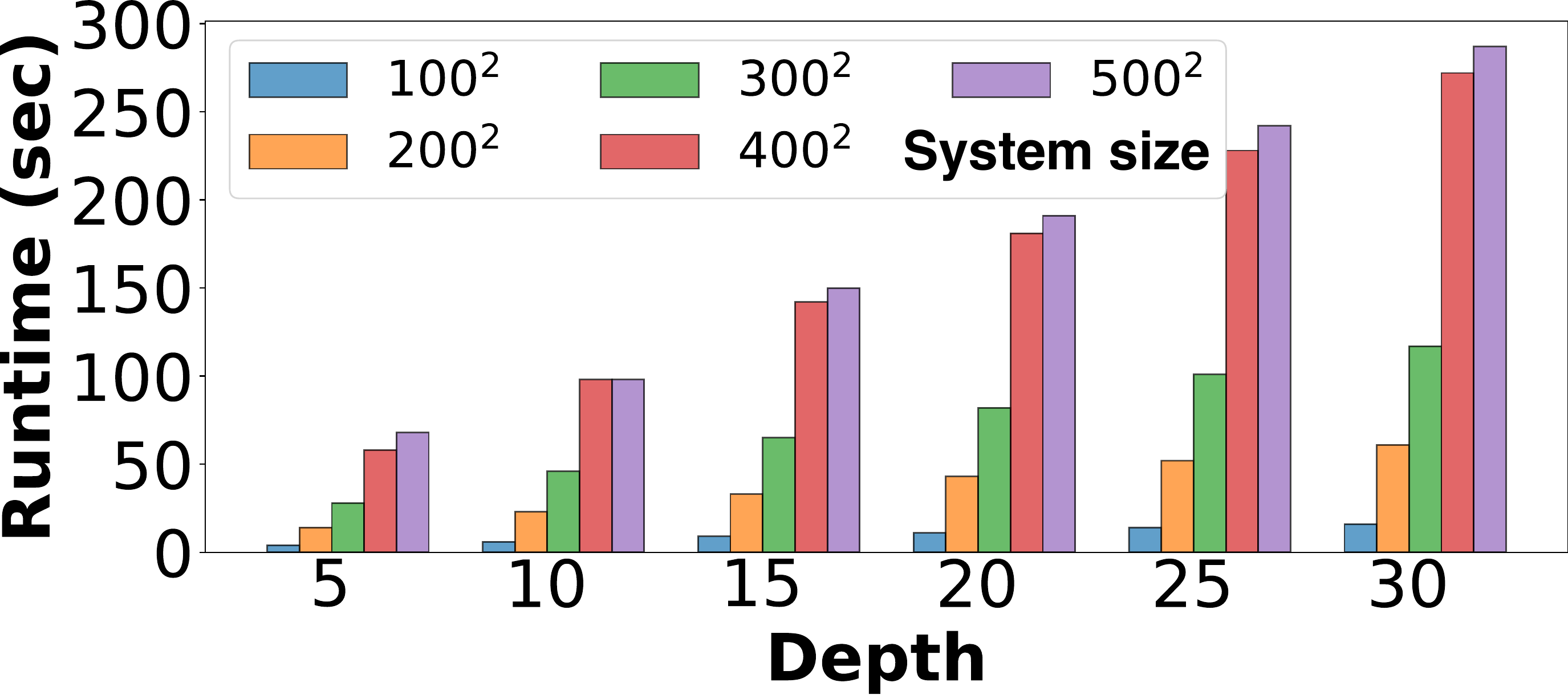}
    \caption{{Large-scale DONNs training runtime.}}
    \label{fig:layers_runtime}
\end{figure}
Furthermore, we evaluate capability of LightRidge of training large DONNs systems (Figure \ref{fig:layers_runtime}). The runtime is acquired on a single Nvidia 3090 Ti GPU. We can see that LightRidge handles 30-layer DONNs training in $\sim$ 280 seconds per epoch, with input image resolution at $500^2$. Besides, we observe runtime increases almost linear w.r.t the DONNs depth, while there is a runtime jump when the system size increases beyond $300^2$, mainly due to the limited resource on a single GPU. This posts strong motivations for further CUDA optimization and multiple-GPU training supports in future works. 
}
{
\begin{table}

\small
\caption{Energy efficiency (fps/Watt) and accuracy comparisons between DONN systems and conventional NNs.}
\label{tbl:energy_comp}
\centering
\begin{tabular}{|c|cc|cc|}
\hline
\multirow{2}{*}{\textbf{Platform}} & \multicolumn{2}{c|}{\textbf{fps/Watt}} & \multicolumn{2}{c|}{\textbf{Accuracy}} \\ \cline{2-5} 
 & \multicolumn{1}{c|}{\textbf{MLP}} & \textbf{CNN} & \multicolumn{1}{l|}{MNIST} & \multicolumn{1}{l|}{FMNIST} \\ \hline
\textbf{GPU 2080 Ti} & \multicolumn{1}{c|}{3.3 (301$\times$)} & 3.8 (261$\times$) & \multicolumn{1}{c|}{\multirow{4}{*}{0.99}} & \multirow{4}{*}{0.91} \\ \cline{1-3}
\textbf{GPU 3090 Ti} & \multicolumn{1}{c|}{2.4 (414$\times$)} & 1.7(585$\times$) & \multicolumn{1}{c|}{} &  \\ \cline{1-3}
\textbf{CPU Xeon} & \multicolumn{1}{c|}{1.5(663$\times$)} & 2.0 (497$\times$) & \multicolumn{1}{c|}{} &  \\ \cline{1-3}
\textbf{XPU (EdgeTPU)} & \multicolumn{1}{c|}{23(43$\times$)} & 26 (38$\times$) & \multicolumn{1}{c|}{} &  \\ \hline
\textbf{\begin{tabular}[c]{@{}c@{}}Our DONNs\\ prototype\end{tabular}} & \multicolumn{2}{c|}{995} & \multicolumn{1}{c|}{0.98} & 0.89 \\ \hline
\end{tabular}
\end{table}
}

{

\subsection{Performance Comparison between DONNs and conventional NNs}
Compared with conventional NN models on digital platforms, the current optical-devices-deployed DONN systems at this early stage suffer from accuracy performance degradation while feature with significantly improved energy efficiency. As shown in Table \ref{tbl:energy_comp}, we evaluate two conventional NNs including a MLP, which consist of two linear layers with hidden size of 128, and the input image is flattened as one-dimensional tensor, i.e., MLP (40000 $\rightarrow$ 128 $\rightarrow$ 10); and a CNN, which consists of two Conv2D, where the kernel size of both layers is set as (5, 5) and 32 filters for the first layer and 64 filters for the second layer with stride and padding being 2, two MaxPooling2D, where kernel size is set as (3, 3) with stride 2, followed by two linear layers. Additionally, we deploy the conventional NNs on different digital platforms including Nvidia GPU 2080 Ti, Nvidia GPU 3090 Ti, Intel Xeon 6230 20x CPU, and Google EdgeTPU \cite{yazdanbakhsh2021evaluation}.

As a result, the conventional NNs can produce the accuracy performance of 0.99/0.99 for MNIST, and 0.91/0.91 for FashionMNIST with the MLP and the CNN model, respectively, while DONN systems reach the accuracy performance of 0.98/0.89 for MNIST/FashionMNIST, which shows 1\% accuracy performance degradation. For practical realization with DONN systems, we take the prototype in Figure \ref{fig:prop_exp} as an example, the power of a CW 532nm laser source is $\sim5$mW. The diffractive layers are passive optical devices and require no extra energy for computation. Then the power consumption at the CMOS detector is $\sim1$ W (max) @ 1000 fps with the system size of $200\times200$. Thus, the power efficiency for the DONN system can be estimated as $995$fps/Watt. The corresponding energy efficiency results for conventional NNs on various digital platforms are shown in Table \ref{tbl:energy_comp}, which shows that the DONN system is roughly 2 orders more efficient than desktop CPU and GPU, and 1 order than digital edge devices with batch size as 1. The energy efficiency provided by DONN systems can be more significant when dealing with more complex ML tasks (e.g., applications in Section \ref{sec:advance_applications}) as the computation part (with passive optical devices) consumes zero power. Note that DONNs energy efficiency can be further optimized with integrated fabrication and high-end detector.

Therefore, the DONN system shows its great potential in completing ML task much more energy-efficiently than conventional NNs. However, the degradation of accuracy performance and the challenges in deploying the practical inference systems call for more future works in broad disciplinaries, such as complex-domain training algorithms, domain-specific co-design, and optics, which also highlights the potential of our framework.

}

\input{isca2022-latex-template/09sec-chip-integration}

\input{isca2022-latex-template/08sec-advanced-arch}

%% file: isca2022-latex-template/09sec-chip-integration.tex
{{
\subsection{On-chip DONNs Integration via LightRidge}\label{sec:integration}

The bulky 3D free-space DONN systems can be integrated as a 3D monolithic on-chip DONNs via 3D additive fabrication \cite{dinc2020computer, dinc2020optical, goi2021nanoprinted, luo2022metasurface}, e.g., galvo-dithered two-photon nanolithography \cite{goi2021nanoprinted}, electron beam lithography overlay process \cite{luo2022metasurface}, etc. Such monolithic on-chip DONNs can be integrated in a hybrid computing system, with DONNs performing as an optical co-processor hosted by central processor via system interconnects (e.g., PCIe 4.0). The host processor controls the laser encoding for loading images and the results collection with the co-processor interconnects, illustrated in Figure \ref{fig:integration}. Each diffractive layer is a thin film, where the trained phase information is encoded with the thickness of the material used for layer fabrications. Between diffractive layers, the optical clear adhesive is employed to provide free-space light diffraction, whose thickness is the diffraction distance. Diffractive layers and optical clear adhesive are stacked sequentially to construct an on-chip DONN system. 
The final prediction is captured on the detector, with Analog-Digital-Converter (ADC), I/O interface, and memory buffers integrated on the peripheral circuits. An example of aforementioned real-world DONN on-chip integration is realized by \cite{luo2022metasurface}. However, due to the three challenges we discussed earlier, the design cycle could take months to year efforts. LightRidge framework can significantly simplify the end-to-end on-chip design process, demonstrated by the case study as follows. 

\noindent
\textbf{Case study} --  We target a 5-layer DONN system integration under wavelength $532$nm for a CMOS detector chip (CS165MU1 from Thorlabs, Inc.), shown in Figure \ref{fig:integration}, where the CMOS chip defines the pixel size of $3.45$um. 
The key for on-chip integration is to search for valid fabrication parameters with high prediction performance w.r.t ML tasks. 
Therefore, following the four steps of LightRidge design flow (Figure \ref{figs:overview}), we first deploy LightRidge-DSE to explore the 3D fabrication dimension, including distance, resolution, and diffraction unit size. {According to the emulation results in Section \ref{sec:archtecture_variation} and Figure \ref{fig:DSE_pixel_dist}(c), when we fix the wavelength as $532$nm and the diffraction unit size (pixel size of the CMOS chip) as $3.45$um, considering image classification as ML tasks (e.g., MNIST), LightRidge-DSE returns the diffraction distance of $532$um, and the resolution 200$\times$200, with the emulation accuracy of $92\%$, to fit the CMOS chip.}
Thus, the DONNs fabrication dimension is finalized as $690\text{um}\times690\text{um}\times2660\text{um}$, where $2660$um is the height, and flat chip dimension is $690\times690\text{um}^2$, which aligns with the chip fabrication procedure in \cite{luo2022metasurface}. Next, after training completed, each layers will be fabricated w.r.t the phase parameters optimized by the codesign stage via nano-printing on the targeted CMOS detector chip. The integrated DONNs can then be used as a co-processor via ADCs and I/O integrated with the CMOS detector chip, where the pre-fabrication design process takes less than a day via LightRidge.

\begin{figure}[!htb]
    \centering
    \includegraphics[width=1\linewidth]{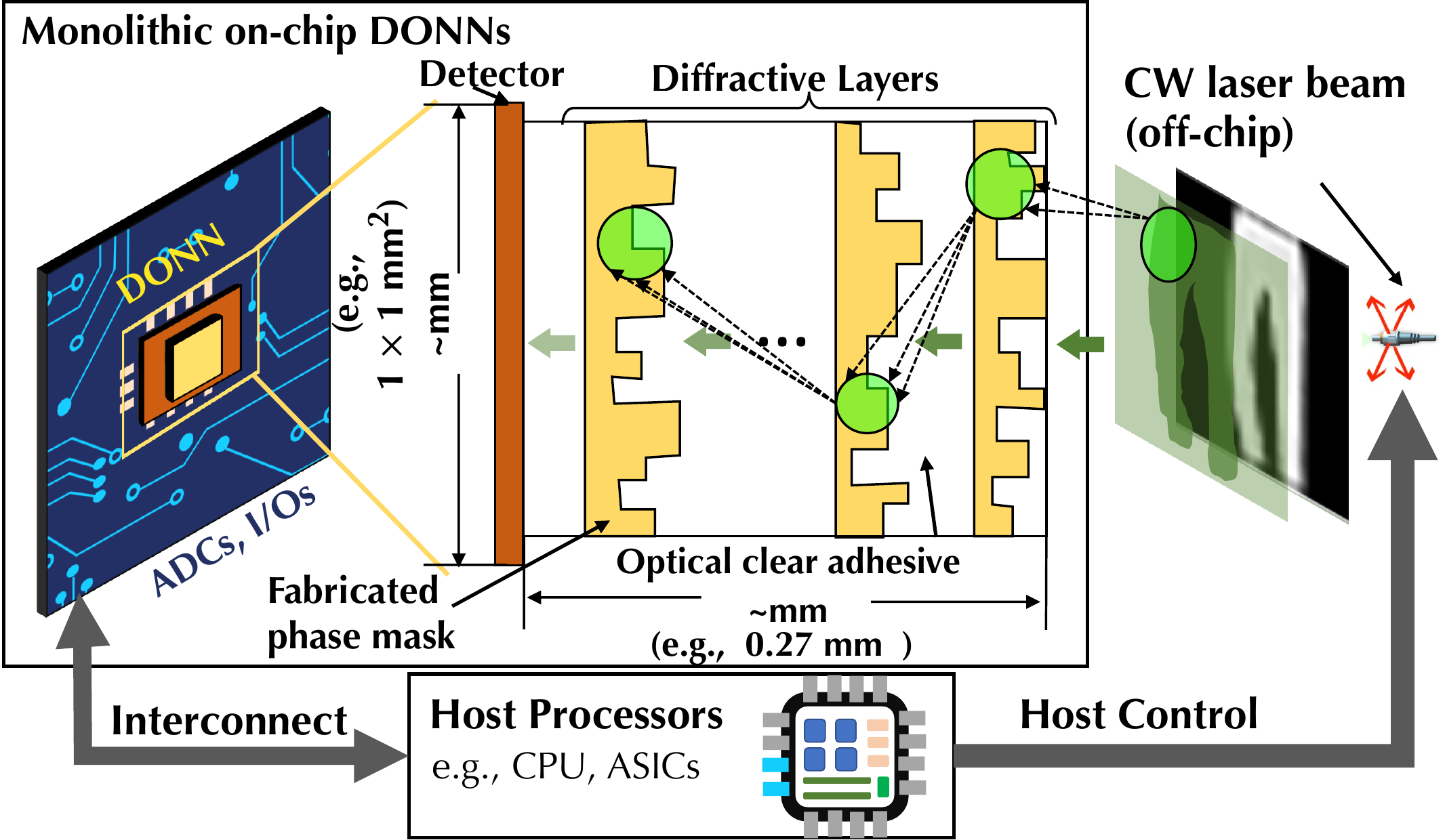}
    \caption{{Monolithic on-chip DONNs design and overall hybrid architecture system integration.}}
    \label{fig:integration}
\end{figure}

}}

%% file: isca2022-latex-template/08sec-advanced-arch.tex
\subsection{{Advanced DONN Architectures}}\label{sec:advance_applications}

With the design capabilities of LightRidge and LightRidge-DSE verified by physical optical systems, we further explore the potentials of DONN systems with more advanced architectures dealing with more complex computer vision tasks. Specifically, we propose and evaluate \textbf{(1)} a multi-channel DONN architecture implemented with diffractive layers to deal with RGB image classifications, and \textbf{(2)} the first all-optical image segmentation demonstration using DONNs with \textit{optical skip connection} for image segmentation and potentially other image-to-image synthesis tasks.

\subsubsection{\textbf{All-optical RGB image classification}}
\label{sec:app_rgb}
 
\begin{figure}[!htb]
     \centering
         \centering \includegraphics[width=0.5\textwidth]{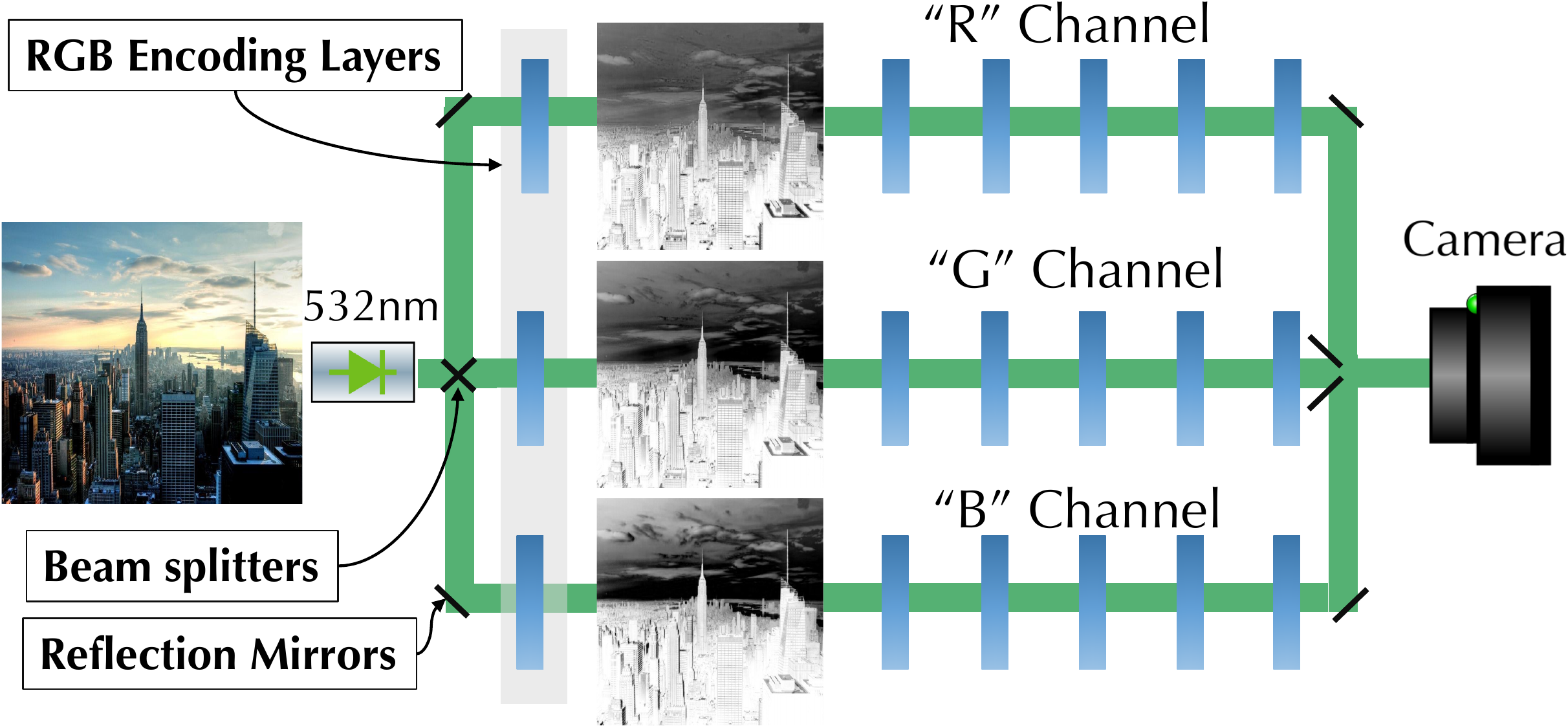}
   \caption{Multi-channel {RGB-DONNs} architecture for image classification using Places365 dataset \cite{zhou2017places}. {The "R/G/B" channels are all encoded as three gray-scaled images using 532 nm laser source.}}
     \label{fig:rgb_arch}
\end{figure}

To deal with more complex datasets in image classification, e.g., Place365 \cite{zhou2017places}, a high-resolution RGB image dataset, we propose a multi-channel RGB-DONNs architecture. As shown in Figure \ref{fig:rgb_arch}, three optical channels are employed in the DONN system to deal with 'R', 'G', 'B' channels separately in the original image, i.e., the original RGB image is split into three 'R'/'G'/'B' channel-only gray-scale images for three optical channels. The input laser beam is split {with the beam splitter into three beams and reflected with mirrors into three channels to encode the corresponding input information}. Note that the image information is encoded with light intensity {at the encoding layer for each channel, in which case each channel takes a gray-scaled image as input and propagates through five diffractive layers. Each channels is constructed with the same system parameters in Section \ref{sec:validation} expect for changing to 5 diffractive layers.} The output laser beams from all channels are projected to a single detector, where the light intensity is merged for the final prediction. Similar to the detector design for classification shown in Figure \ref{fig:onn_full}, a single detector collects the intensity of the output within each pre-defined detector region and produces the predicted class by \texttt{argmax}. All three channels are trained w.r.t the same shared loss function. 

\begin{table}[!htb]
\centering
\begin{tabular}{|c|c|c|c|}
\hline
\textbf{Places365\cite{zhou2017places}} & \textit{\textbf{Top-1}} & \textit{\textbf{Top-3}} & \textit{\textbf{Top-5}} \\ \hline
Our (Fig. \ref{fig:rgb_arch}) & \textbf{0.52} & \textbf{0.73}& \textbf{0.84}\\ \hline
Baseline \cite{zhou2021large} & 0.23 & 0.48 & 0.67\\ \hline
\end{tabular}
\caption{Classification accuracy on Places365 (standard, 256-by-256) with \textit{type of environment} as classes.}
\label{tbl:places365}
\end{table}

The {emulation} accuracy results for image classification with Place365 are shown in Table \ref{tbl:places365}, including top-1, top-3, and top-5 accuracy. The baseline is the {emulation} accuracy from the DONN model trained with the algorithm in \cite{zhou2021large}. The model trained with our framework has better accuracy performance than the baseline in all accuracy matrix ($29\%$/$25\%$/$17\%$ improvement for top-1/top-3/top-5 accuracy, respectively), and ours outperforms the baseline most at the top-1 accuracy. 

\begin{figure}[!htb]
     \centering
      \begin{subfigure}[b]{0.5\textwidth}
         \centering
    \includegraphics[width=0.99\textwidth]{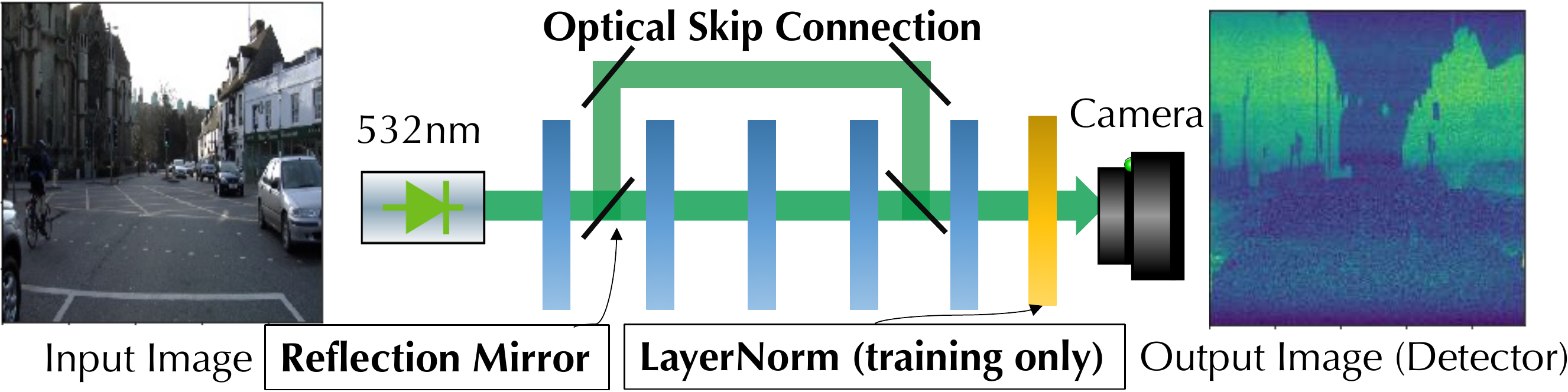}
         \caption{DONNs architecture for image segmentation tasks with optical skip connections and layer normalization (only for training process).}
        \label{fig:seg_1}
     \end{subfigure}
           \begin{subfigure}[b]{0.5\textwidth}
         \centering
    \includegraphics[width=1\textwidth]{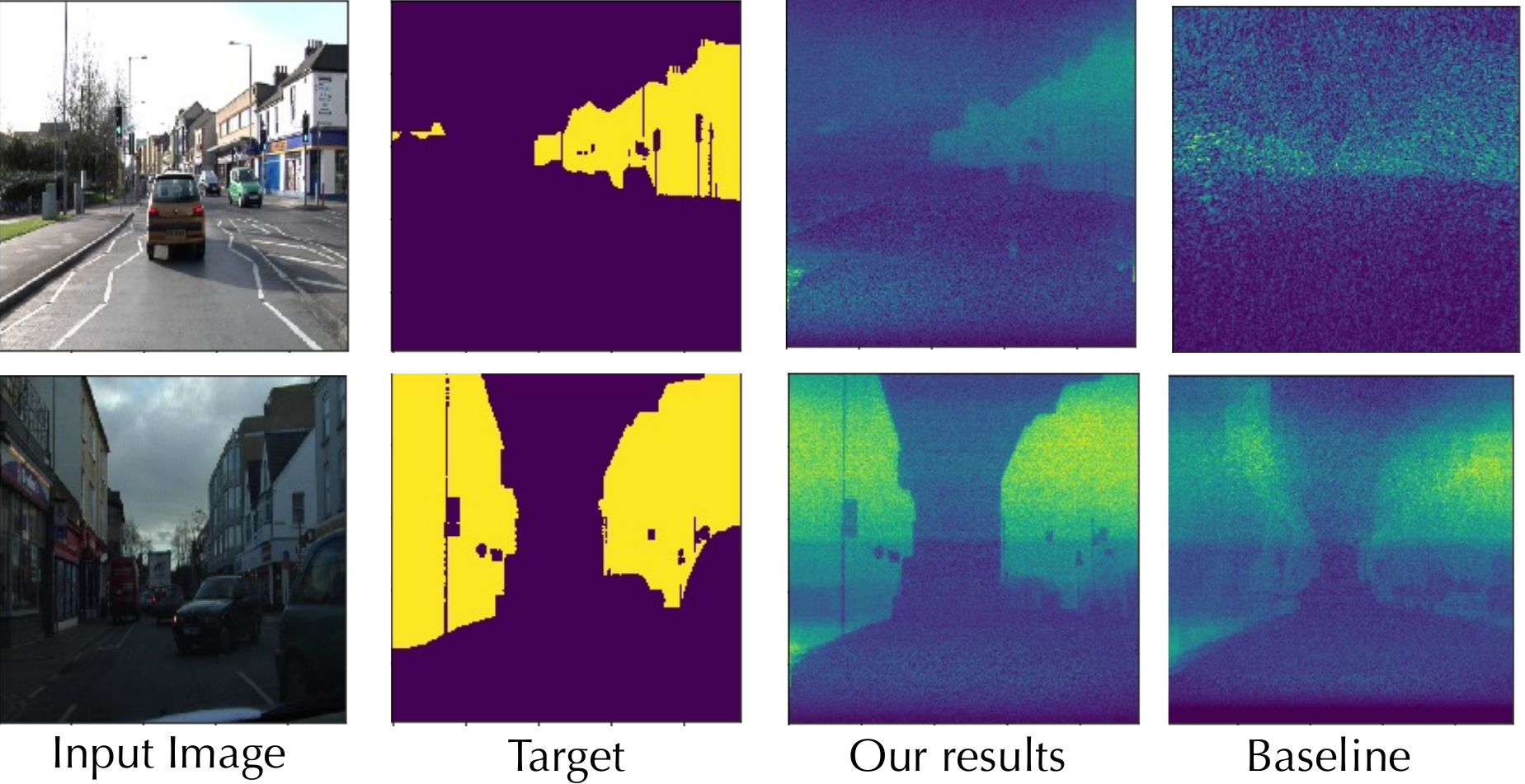}
         \caption{LightRidge enabled advanced segmentation DONNs compared to baseline models and training methods proposed in \cite{lin2018all,zhou2021large}.}
        \label{fig:seg_2}
     \end{subfigure}
 
     \caption{Image segmentation demonstrations using CityScapes \cite{cordts2016cityscapes} datasets with \textbf{(a)} a novel advanced DONNs architecture with optical skip connection and layer normalization for improving training efficiency, and \textbf{(b)} evaluations and comparisons to SOTA baselines \cite{lin2018all,zhou2021large}.}
     \label{fig:seg}
\end{figure}

\subsubsection{\textbf{All-Optical image segmentation}}
\label{sec:app_seg}

Image segmentation is an important and challenging task in modern computer vision tasks, which has a great impact on autonomous systems such as autonomous driving, robotics, etc. Unlike image classification tasks, image segmentation is a process of generating representations of an image into specific image-to-image objectives. While in DONN classification systems, we observe that the system (output detector in particular) is not fully utilized, as only a given number of small detector regions are used for classification. As the DONN system propagates the input image w.r.t trained phase modulations in the full spatial dimension of the system, it is expected to be able to deal with image-to-image based tasks. Thus, we design and demonstrate the first-ever all-optical image segmentation.

Figure \ref{fig:seg_1} includes the proposed 5-layer DONN system, where we introduce two innovations in DONN architecture: 1) optical skip connection, {which is inspired from the residual block design in conventional ResNet \cite{he2016deep} architecture. It aims to smooth the gradient for better training performance and also is involved in inference for better detailed segmentation. Since the light signal is aggressively diffracted during the propagation, the optical skip connection can help to restore some features from less-diffracted inputs, making the model prediction be aware of the original information, which is confirmed to introduce better image segmentation performance with our results;} and 2) layer normalization \cite{ba2016layer} before the detector plane, {which is only employed in the training process for better training performance of the DONN by smoothing the training gradients.} 
The dataset we demonstrate here is selected from CityScapes dataset \cite{cordts2016cityscapes}, where the images are {converted to gray-scaled images and }resized to $350 \times 350$. We use binary labels in this case study to generate segmentation masks for \textit{buildings} and others. The baseline is the results from the DONN model construction without optical skip connection and the training method without layer normalization proposed in \cite{lin2018all, zhou2021large}. {The system parameters and training setups are the same as discussed in Section \ref{sec:validation} expect for the system size changing to $350 \times 350$ and the model structure changing to Figure \ref{fig:seg_1}. }The results shown in Figure \ref{fig:seg_2} demonstrate that the advanced model trained with LightRidge outperforms the baseline in edge detection and with significant clarity improvements on small objects segmentation. These advanced DONN architectures and validations demonstrate the generalizability and power of LightRidge in exploring new architectural designs and applications.

%% file: isca2022-latex-template/10sec-conclusion.tex
\section{Conclusion and Future Work}
\label{sec:conclusion}

This work presents an agile end-to-end design framework LightRidge that enables seamless design-to-deployment of DONNs. LightRidge accelerates and simplifies the design, exploration, and on-chip integration by offering highly versatile and runtime efficient programming modules, and DSE (LightRidge-DSE) engine to construct and train the DONN systems in a wide range of optical settings. The high-performance physics emulation kernels are optimized for runtime efficiency, and verified together with hardware-software codesign algorithm on our visible-range prototype.
Additionally, two advanced DONN architectural designs constructed with LightRidge show the capabilities and generalizability of LightRidge for various ML tasks and system design explorations. We believe our framework LightRidge will enable collaborative and open-source research in optical accelerators for not only ML tasks, but also other optics-related research areas such as optical structure emulation, chip fabrication (lithography), meta-material exploration, etc. 

{In the future, we will further optimize the runtime efficiency of LightRidge, including realizing high-performance CUDA kernel optimization and multiple-GPU computation. }
Additionally, as we have initialized full-chip integration (Figure \ref{fig:integration}) in embedded SoC system, we can deploy our advanced DONNs in image segmentation enabled by LightRidge to demonstrate first all-optical autonomous driving prototype.
We also expect more functionality to be integrated in the framework and more hardware prototypes for experimental demonstrations. For example, the non-linearity in DONN systems, which can be realized by nonlinear optical materials (crystals, polymers, graphane, etc.), is an important implementation for more complex DONNs systems. Finally, we can employ LightRidge for optical phenomena exploration such as the interpixel crosstalks in optical field, which happens when there is a sharp phase change between adjacent phase modulators~\cite{lou2023effects, zhou2023physics}.

%% file: isca2022-latex-template/ae.tex
%
%
%
%
%






\appendix
\section{Artifact Appendix}

\subsection{Abstract}

LightRidge is an open-source framework for end-to-end optical machine learning (ML) compilation, which connects physics to system. This framework provides user-friendly DSL to design, explore and deploy the DONN system with customization. To use the framework, the implemented computation kernels are enclosed within a Python package with easy installation of \textit{pip install lightridge}. Two execution files for raw DONN emulation and the hardware-aware codesign DONN emulation are provided in \textit{tutorial\_01\_raw.py} and \textit{tutorial\_01\_codesign.py}, respectively. We also provide the bash script \textit{run.sh} for the exploration flow including model training, model inference, and model visualization (See \href{https://github.com/lightridge/lightridge/tree/main/ASPLOS2024_AE}{Python tutorial}). Furthermore, we provide Colab tutorial for easy interactive training and visualization access (See \href{https://drive.google.com/file/d/1DVzvvei0GBXX_jZewNtusR7xLCA-TwDM/view?usp=sharing}{Google Colab Access}.).

\subsection{Artifact check-list (meta-information)}

{\small
\begin{itemize}
  \item {\bf Program:} \texttt{Python}
  \item {\bf Model: } Diffraction optical neural networks.
  \item {\bf Data set:} Public data included in examples. Dataloaders are provided to elaborate additional datasets.
  \item {\bf Run-time environment:} Ubuntu 18 or above, macOS 10.15 (Catalina) or above, and RHEL.
  \item {\bf Hardware: } Minimum 256 GB storage, 64 GB memory, with CPU and GPU and compatible driver/library.
  \item {\bf Output: } Diffractive optical neural network numerical models, performance metrics, and visualization.
  \item {\bf Experiments: } The training, inference, and visualization of the 5-layer DONN system for the image classification task with MNIST10~\cite{lecun1998mnist} dataset. 
  \item {\bf How much time is needed to prepare workflow (approximately)?: } Less than 10 min.
  \item {\bf Publicly available?: } Yes. ''\texttt{pip install lightridge}'', and Google Colab Tutorials available.
  \item {\bf Code licenses: } GNU GPL 3.0
\end{itemize}
}

\subsection{Description}

\subsubsection{How to access}

LightRidge is packaged and publicly released at The Python Package Index (PyPI). The package is available to access and install by \texttt{pip install lightridge}.

\begin{itemize}
    \item Colab tutorial:  \href{https://drive.google.com/file/d/1DVzvvei0GBXX_jZewNtusR7xLCA-TwDM/view?usp=sharing}{Google Colab Access}. 
    \item Bash/Python scription tutorials: \url{https://github.com/lightridge/lightridge/tree/main/ASPLOS2024_AE}
    \item Visit LightRidge website at \url{https://lightridge.github.io/lightridge} for additional tutorials and documentations of the infrastructure. \href{https://lightridge.github.io/lightridge/lightridge_tutorial_ASPLOS24_AE.html}{Access the specific AE page}.
\end{itemize}


\subsubsection{Hardware dependencies}

The accuracy performance can vary on different GPU versions with limited training efforts for demonstration. However, when trained with enough training efforts, e.g., enough training epochs, the difference between different hardware platforms is negligible. 

We provide two platforms for demonstration:

\begin{itemize}
    \item \textbf{Colab} for easy access and demonstration -- This tutorial is deployed on the T4 GPU of Colab, which is the default GPU as free GPU resources. In case of fast and feasible runtime in Colab, we reduce the training efforts to only 5 epochs for the raw DONN emulation and 3 epochs for the codesign DONN emulation, which results in degraded performance in Colab tutorial compared with the claimed performance in the paper. \href{https://drive.google.com/file/d/1DVzvvei0GBXX_jZewNtusR7xLCA-TwDM/view?usp=sharing}{Google Colab Access}.
    
    \item \textbf{Python Scripts} deployed on the server with dedicated Nvidia GPU (CUDA$\geq$11.x) -- These python files are implemented for command run on the server. Our implementations have been evaluated with 2080 Ti, 3090, and 4090 GPUs, with Intel Xeon(R) Gold 6230 CPU.
\end{itemize}

\subsubsection{Data sets}

We provide MNIST10~\cite{lecun1998mnist} dataset for demonstration. We also provide the parameter in the code to download FashionMNIST~\cite{xiao2017fashion} for image classification task. Moreover, we provide customizable dataloader in \texttt{lr.utils} for loading your own datasets.

\subsubsection{Models}

In this tutorial, we configure the model as a 5-layer DONN model with the same system setups shown in Section \ref{sec:validation} for demonstration. We provide both pre-configured basic DONN models in the python package in \texttt{lr.models} and the customized model construction demonstration in the main execution python file. Both are constructing a sequentially stacked 5-layer DONN systems for demonstration.

Following our comments in the Colab tutorial, you will be able to configure the optical neural architecture by modifying the parameters such as:

\begin{itemize}
        \item Laser source information: \texttt{wavelength} (default: 532e-9 in meter)
        \item System resolution/size: \texttt{sys\_size} (200)
        \item Pixel dimension: \texttt{pixel\_size} (3.6e-5 in meter)
        \item Mathematical approximation for light diffraction: \texttt{approx}
        \item Diffraction distance: \texttt{distance} (default: 0.3 in meter).
        \item System depth: \texttt{num\_layers} (default: 5)
\end{itemize}

\subsection{Installation}

Released as Packaged Python Project \texttt{lightridge}: \texttt{pip install lightridge}. Other two main execution files \textit{tutorial\_01\_raw.py} and \textit{tutorial\_02\_codesign.py}, and a bash script \textit{run.sh} are included in the folder.

\subsection{Experiment workflow}

We provide two approaches for the demonstration:
\begin{itemize}
    \item For Colab tutorial, the code block is run one by one for interactive results feedback.

    \item For the python files, the execution flow for model training, inference, and visualization is implemented in \textit{run.sh}.
\end{itemize}

\noindent
The detailed experimental pipeline are organized as follows:

\begin{itemize}
    \item Step 1: LightRidge installation
    \item Step 2: Check LightRidge installation
    \item Step 3: Load packages and configure training devices
    \item Step 4: Constructing DONNs
    \item Step 5: Training DONNs
    \item Step 6: Visualization
    \item Step 7: Change the DONN model with codesign information
    \item Step 8: Add the device parameters
    \item Step 9: Visualization
\end{itemize}

\subsection{Evaluation and expected results}

Expected results should match \url{https://github.com/lightridge/lightridge/tree/main/ASPLOS2024_AE} and \href{https://lightridge.github.io/lightridge/lightridge_tutorial_ASPLOS24_AE.html}{Access the specific AE page}, in 1) accuracy metrics and 2) propagation and phase visualization.

\subsection{Experiment customization}

This framework provides customization for both model constructions and training setups. Different model constructions involve exploration efforts to find the paired parameters as shown in Section~\ref{sec:archtecture_variation}. The training setups such as learning rate (\textit{--lr}), training epochs (\textit{--epochs}), batch size (\textit{--batch\_size}), etc., can be customized by the arguments implemented in the python file.


